\documentclass[superscriptaddress, nofootinbib, twocolumn, amsmath,amssymb, aps, pra, notitlepage, longbibliography]{revtex4-1}

\usepackage{dcolumn}
\usepackage{bm}
\usepackage{epsfig}
\usepackage{graphicx}
\usepackage{latexsym}
\usepackage{amsfonts}
\usepackage{setspace}
\usepackage{graphicx}
\usepackage{amsmath}
\usepackage{verbatim}
\usepackage{color}
\usepackage{SIunits}
\usepackage{hyperref}
\usepackage{nccmath}
\usepackage{upgreek}

\begin{document}

\title{On-chip broadband non-reciprocal light storage}

\author{Moritz Merklein$^{1,2,\ast,\dagger}$, Birgit Stiller$^{1,2,3\ast}$, Khu Vu$^{3}$,  Pan Ma$^{3}$, Stephen J. Madden$^{3}$, and Benjamin J. Eggleton$^{1,2}$\\
\small{ \textcolor{white}{blanc\\}
$^{1}$The University of Sydney Nano Institute (Sydney Nano), The University of Sydney, NSW~2006, Australia.\\
$^{2}$Institute of Photonics and Optical Science (IPOS), School of Physics, The University of Sydney, NSW 2006, Australia.\\
$^{3}$Max-Planck-Institute for the Science of Light, Staudtstr. 2, 91058 Erlangen, Germany.\\
$^{4}$Laser Physics Centre, Research School of Physics and Engineering, Australian National University, Canberra, ACT 2601, Australia.\\
$^{\ast}$These authors contributed equally to this work.\\
$^{\dagger}$moritz.merklein@sydney.edu.au}}

\begin{abstract} 

Breaking the symmetry between forward and backward propagating optical modes is of fundamental scientific interest and enables crucial functionalities, such as isolators, circulators, and duplex communication systems. Whereas there has been progress in achieving optical isolation on-chip, integrated broadband non-reciprocal signal processing functionalities that enable transmitting and receiving via the same low-loss planar waveguide, without altering the frequency or mode of the signal, remain elusive.
Here, we demonstrate a non-reciprocal delay scheme based on the uni-directional transfer of optical data pulses to acoustic waves in a chip-based integration platform. We experimentally demonstrate that this scheme is not impacted by simultaneously counter-propagating optical signals. Furthermore, we achieve a bandwidth more than an order of magnitude broader than the intrinsic opto-acoustic linewidth, linear operation for a wide range of signal powers, and importantly, show that this scheme is wavelength preserving and avoids complicated multi-mode structures.

\end{abstract}

\maketitle


\section{Introduction}
Reciprocity is a general concept in optics dictating that a transmission channel does not change, or is symmetric, under the interchange of source and receiver \cite{Jalas2013,2017,Caloz2018}. There are, however, different approaches to break this reciprocity, most commonly by utilizing magnetic materials \cite{Shoji2008,Bi2011,Huang2017a}. The arguably most common devices based on breaking reciprocity are isolators utilizing magnetic materials. Isolators based on magnetic materials are passive components and have large bandwidth and rejection. Chip integration and material losses, however, are still major challenges despite the great progress made over the recent years. \newline
Another way to achieve non-reciprocal transmission is based on temporal modulation \cite{2017,Yu2009}. While this active method requires some sort of pumping - electrical \cite{Lira2012}, optical \cite{Kang2011}, or acoustic \cite{Sohn2018} - it offers great potential as it does not rely on magnetic materials and hence is more suitable for chip integration with the additional advantage of being reconfigurable. Non-reciprocal elements, such as optical isolators and circulators are crucial building blocks, in particular, for integrated photonic circuits to protect the laser but also to route counter-propagating signals and mitigate back-reflections that can arise from boundaries between multiple elements or are simply caused by Rayleigh backscattering. \newline
A powerful and versatile way to achieve non-reciprocal transmission in a small footprint can be realized by coupling light and mechanical degrees of freedom \cite{Verhagen2017,Miri2017}. This coupling between light and mechanical oscillations is greatly enhanced in resonant structures which lead to demonstrations of chip-scale optomechanical isolators and circulators for optical \cite{Achar2008,Verhagen2012a,Xu2015b,Ruesink2016,Shen2016,Fang2016b,Shen2018} as well as microwave signals \cite{Estep2016,Peterson2017b,Barzanjeh2017,Bernier2017}. \newline
Similarly, stimulated Brillouin scattering (SBS), i.e. the resonant coupling of optical waves with propagating acoustic waves in waveguides via optically induced forces is known to be non-reciprocal \cite{Kang2011,Huang2011}. The phase-matching condition ensures that only pump and probe waves which are counter-propagating (co-propagating) couple to the acoustic wave which is mainly longitudinal for backward SBS (transverse for forward Brillouin scattering) \cite{Eggleton2019}. As these acoustic waves carry momentum, Brillouin interactions can be used to induce indirect photonic transitions between different optical modes \cite{Kang2011,Huang2011,Poulton2012c}. \newline
Numerous experiments have reported non-reciprocity exploiting Brillouin interactions, including demonstrations in photonic crystal fibers \cite{Kang2011}, silicon waveguides \cite{Kittlaus2018}, and fiber-tip resonators \cite{Kim2015, Dong2015, Kim2017a}. Surprisingly, however, so far there was no demonstration of Brillouin-based non-reciprocal schemes harnessing backward SBS. In this approach, the optical wave can couple to a continuum of acoustic modes that provides enormous flexibility, which is particularly important for non-reciprocal signal processing schemes that go beyond providing pure signal isolation. Furthermore, large backward SBS gain can be achieved in small footprint planar integrated circuits \cite{Eggleton2019}. \newline
Here, we show a non-reciprocal light storage scheme based on coherent Brillouin coupling of acoustic and optical modes to achieve non-reciprocal delay. We experimentally demonstrate that optical pulses that are traveling simultaneously in the opposite direction, with the same optical frequency and mode, are not impacted nor do they impact the storage process. We show that the bandwidth of the scheme can be broadened beyond the intrinsic acoustic linewidth - in this demonstration by more than one order of magnitude - which was generally thought to be a limiting factor of optomechanical non-reciprocal schemes. Furthermore, the scheme depends linearly on the input data pulse power in the observed range and does neither alter the frequency nor mode of the incoming data - all important requirements for practical non-reciprocal devices. \newline
\section{Results}
\begin{figure}[h!]
\begin{center}
  \includegraphics[width=0.90\linewidth]{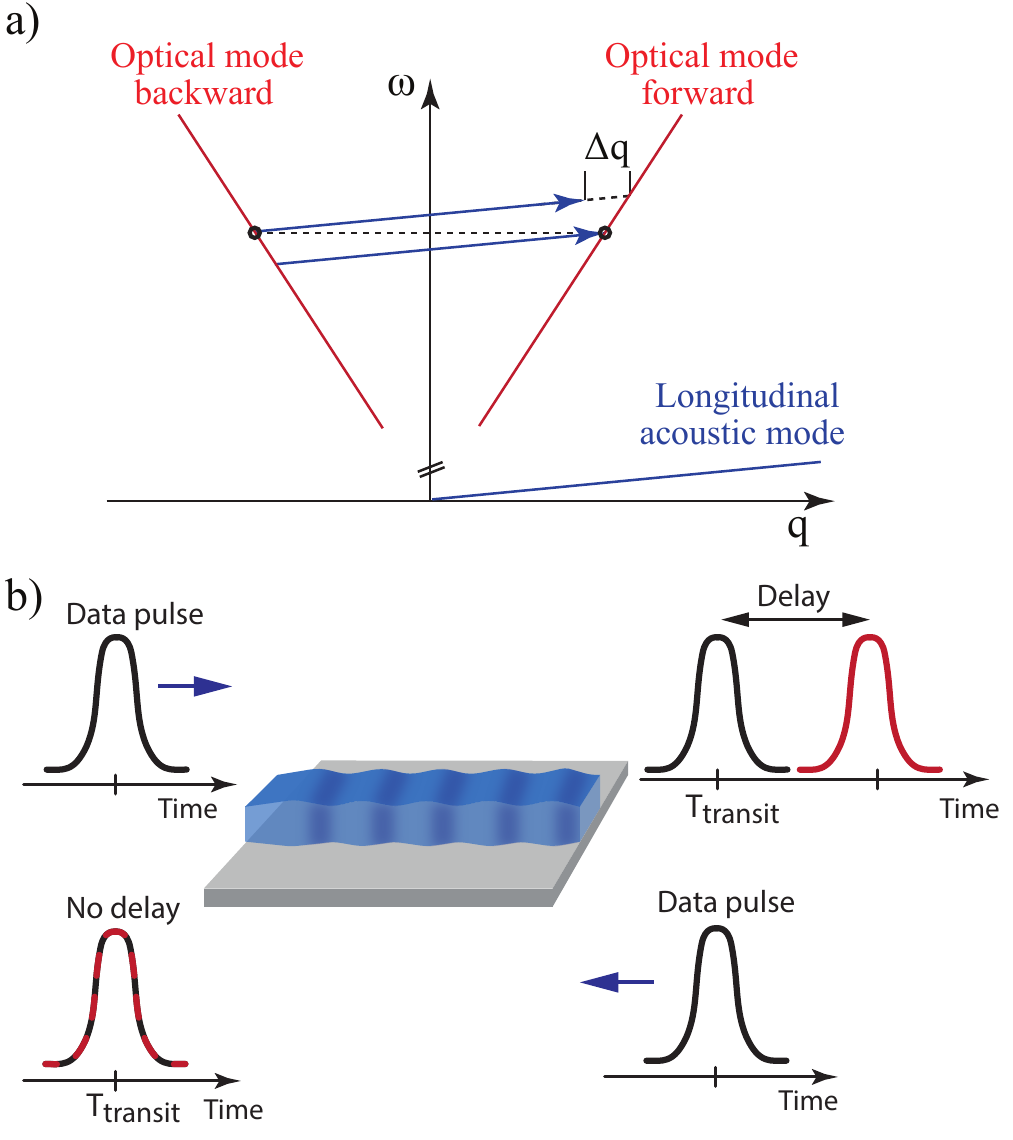}
  \caption{Basic principle and phase-matching diagram. a) Optical data pulses that are coupled from the left side into the waveguide are converted to acoustic phonons and experience a delay. Optical data pulses that are coupled simultaneously from the opposite side in the waveguide do not experience any conversion and hence are not delayed. b) Dispersion diagram for backward Brillouin scattering illustrating the phase-matching condition for data pulses propagating in opposite directions.}
\label{princ}
\end{center}
\end{figure}

The non-reciprocity is induced by an interaction between two counter-propagating optical modes - here we call them data \(\omega_\mathrm{data}\) and write/\,read \(\omega_\mathrm{w/r}\) - with a traveling acoustic mode \(\Omega\). This interaction has not only to fulfill energy conservation \(\omega_\mathrm{data} = \omega_\mathrm{w/r} + \Omega\) but also momentum conservation \(\mathbf{k}_\mathrm{data} = \mathbf{k}_\mathrm{w/r} + \mathbf{q}\). As the momentum \(\mathbf{q}\) of the traveling acoustic wave in backward SBS is large, approximately twice the magnitude of the momentum vectors of the individual optical modes \(|\mathbf{q}| \approx 2 \cdot |\mathbf{k}_\mathrm{data}|\) with \(|\mathbf{k}_\mathrm{data}| \approx |\mathbf{k}_\mathrm{w/r}|\), the interaction is only phase-matched in one direction. This becomes more evident when looking at the dispersion diagram shown in Fig. \ref{princ}\,a, where changing the direction of one optical mode leads to a phase mismatch \(\Delta \mathbf{q}\). As the interaction takes place over an elongated length given by either the length of the waveguide, for the case of continuous wave (CW) signals, or the length of the optical signal pulse, a small initial phase mismatch builds up to a large mismatch over that interaction length as it was shown for the case of Brillouin interactions that involve multiple optical wavelengths \cite{Stiller2019}. Even in the case of nanosecond (ns) and sub-ns pulses this length scale is in the order of several centimeters. \newline
How this strict phase-matching condition of the opto-acoustic Brillouin interaction can be utilized to achieve uni-directional signal delays is shown in Fig. \ref{princ}\,b. Optical data pulses \(\omega_\mathrm{data}\) that propagate from the left through the waveguide are transferred to acoustic phonons via a Brillouin interaction induced by counter-propagating write pulses \(\omega_\mathrm{w}\) and are subsequently retrieved using read pulses \(\omega_\mathrm{r}\) \cite{Merklein2017,Zhu2007,Merklein2018} whereas optical pulses that travel in the opposite direction are neither effected by the write, the read pulses nor the acoustic wave that stores the original optical pulses \(\omega_\mathrm{data}\). The phase matching condition between the optical modes and the traveling acoustic wave ensures that there is no interaction, even for the extreme case when the counter-propagating optical pulses are in the same optical waveguide mode, at the same frequency and optical power. \newline
Data pulses that are transferred to the acoustic domain accumulate a large delay due to the five orders of magnitude difference in velocity between the acoustic and optical waves. The accumulated delay of the optical signal can hence be approximated by the time difference between the writing and the retrieving operation. Thus the delay of the optical signal can be continuously tuned within the acoustic lifetime of the acoustic phonon that is given by the material properties of the chalcogenide glass and is in the order of 10\,ns \cite{Merklein2017}.  \newline
To enable efficient coupling from optical to acoustic waves, the optical write/\,read and the data pulses are separated by the frequency of the acoustic wave in the waveguide \(\Omega = 2 \cdot n_\mathrm{eff} \cdot v_\mathrm{ac} \cdot \lambda^{-1}\), where \(n_\mathrm{eff}\) is the effective refractive index, \(v_\mathrm{ac}\) the acoustic sound velocity in the material and \(\lambda\) the optical pump wavelength. The waveguide used in this demonstration is made out of chalcogenide glass and the corresponding acoustic resonance frequency is \(\Omega \approx 7.6\,\mathrm{GHz}\). \newline
Due to the phase-matching condition of the Brillouin interaction the coupling between optical and acoustic wave only occurs for a certain pair of optical write/\,read and data pulses. Optical data pulses that are simultaneously propagating through the waveguide from the opposite side are neither transferred to the acoustic wave by the write pulses, nor do they interact with the acoustic wave present in the waveguide and hence do also not influence the stored data pulses. The counter-propagating signals could be data that is simultaneously transmitted/\,received in the opposite direction through the same waveguide or could simply originate from back-scattering in the photonic circuit. \newline
\begin{figure*}[t]
\begin{center}
  \includegraphics[width=0.67\textwidth]{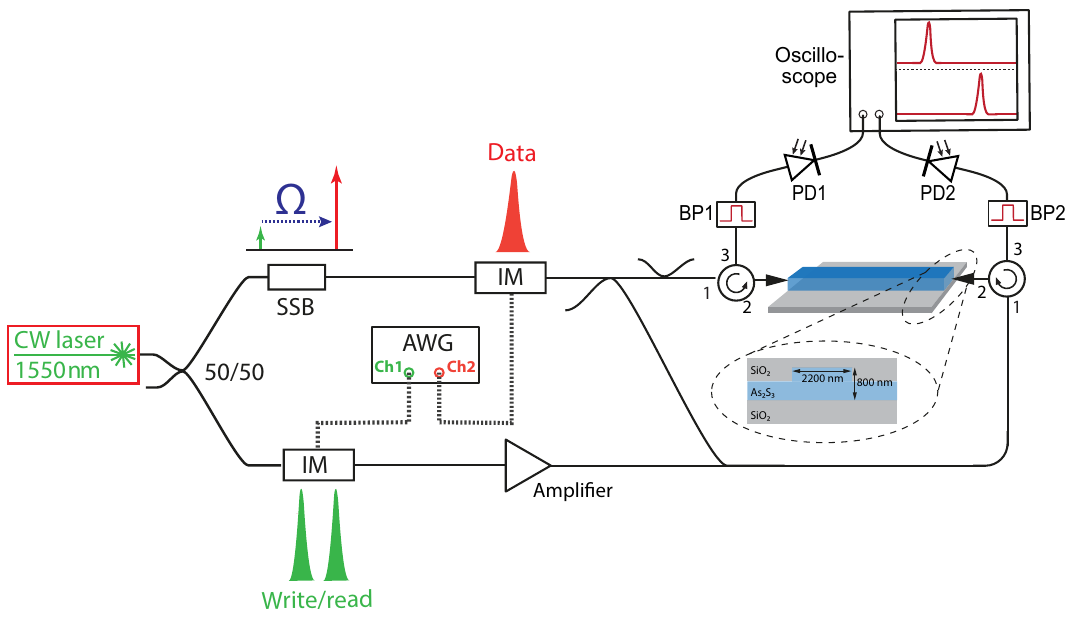}
  \caption{Schematic experimental setup. CW laser: continuous wave distributed feedback (DFB) laser; 50/50 fiber coupler; SSB: single-sideband modulator; IM: intensity modulator; AWG: multi-channel arbitrary waveform generator; Amplifier: Erbium-doped fiber amplifier;  BP: Bandpass filter; PD: photodetector. Inset: cross-section of the chalcogenide rib waveguide embedded in silica.}
\label{setup}
\end{center}{}
\end{figure*}
\subsection{Experimental setup}
The simplified experimental setup is shown in Fig. \ref{setup} (a detailed scheme and description can be found in the Materials and Methods section \ref{methods}). A continuous-wave (CW) distributed feedback laser is split into two paths, the data and the write/\,read path. The data signal is shifted by the Brillouin frequency shift \(\Omega\) of the waveguide. Afterwards, the CW signals in both arms are modulated into short pulses using a multi-channel arbitrary waveform generator and electro-optic intensity modulators. The data signal is split using a 50/50 coupler. One part of the signal is combined with the write/\,read arm and coupled to the nonlinear chalcogenide waveguide. The other half passes through an additional 50/50 coupler, to ensure that the path length of both data signals as well as their optical power is the same in both arms, and is coupled to the chalcogenide chip from the opposite side. On both sides of the chip circulators are used to separate in- and output. Two narrowband filters (bandwidth \(\approx 3\)\,GHz) are used to separate the data signal from the pump signal and fast photodetectors (bandwidth \(>10\)\,GHz) and a fast oscilloscope (bandwidth \(>10\)\,GHz) are used to detect the data pulses. A cross-section of the chalcogenide waveguides is shown in the inset of Fig. \ref{setup}. A rib structure with a cross-section of 2200 \(\times\) 850\,nm\(^2\) is used to reduce losses caused by sidewall roughness. The chalcogenide glass is surrounded by silica glass to ensure confinement of the optical as well as the acoustic mode \cite{Poulton2013a}. For details on the fabrication of the acousto-optic waveguides see the Materials and Methods section. \newline
\subsection{Non-reciprocal Brillouin light storage}
\begin{figure}[b]
\centering
  \includegraphics[width=0.7\linewidth]{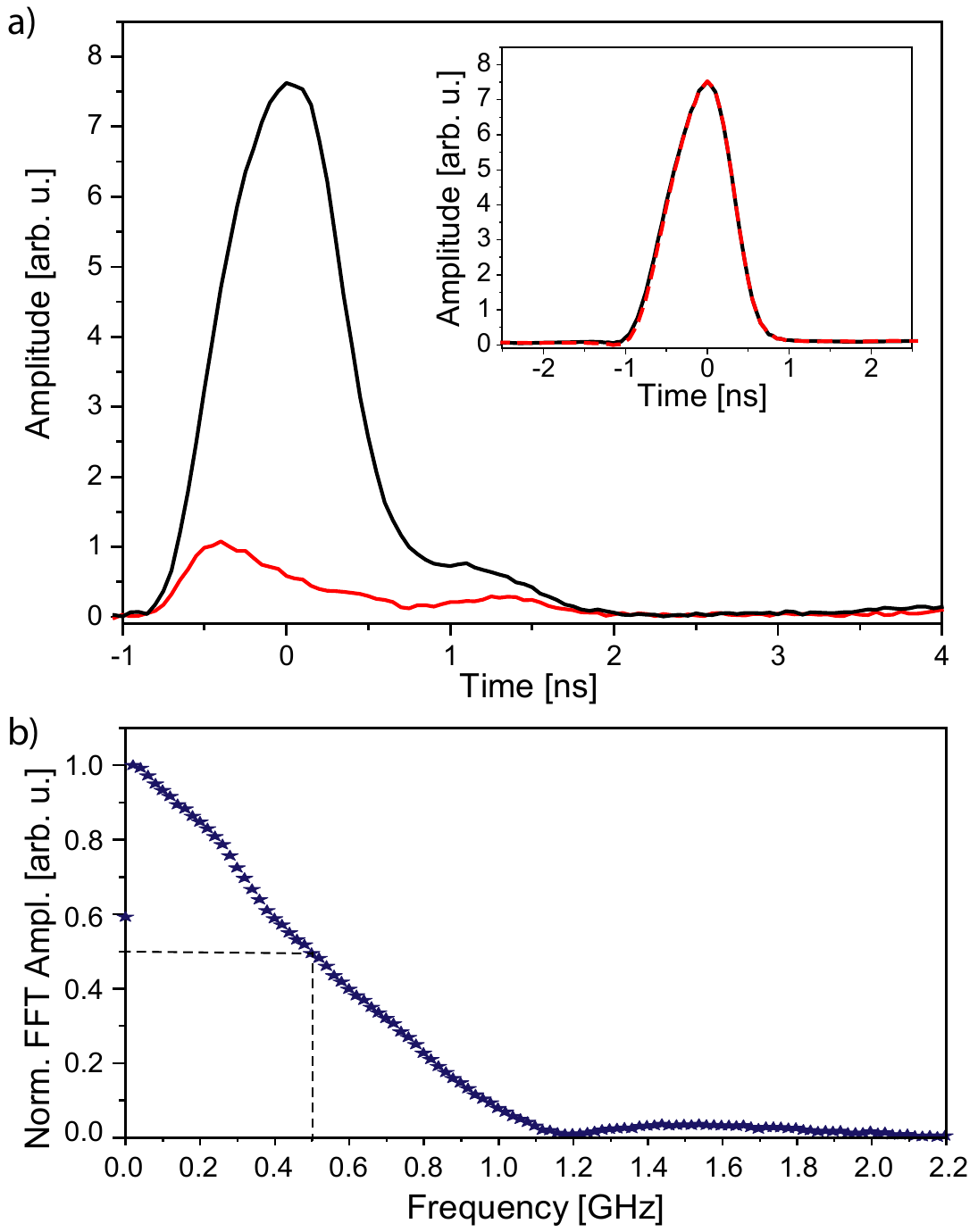}
  \caption{Non-reciprocal pulse depletion. a) Conversion of an optical data pulse (black curve) to acoustic phonons (red curve) while a simultaneously counter-propagating data pulse is not impacted (black and dashed red curve in inset). b) Fast Fourier transform (FFT) of the input data pulse.}
\label{depl}
\end{figure}
\begin{figure*}[t]
\begin{center}
  \includegraphics[width=0.62\textwidth]{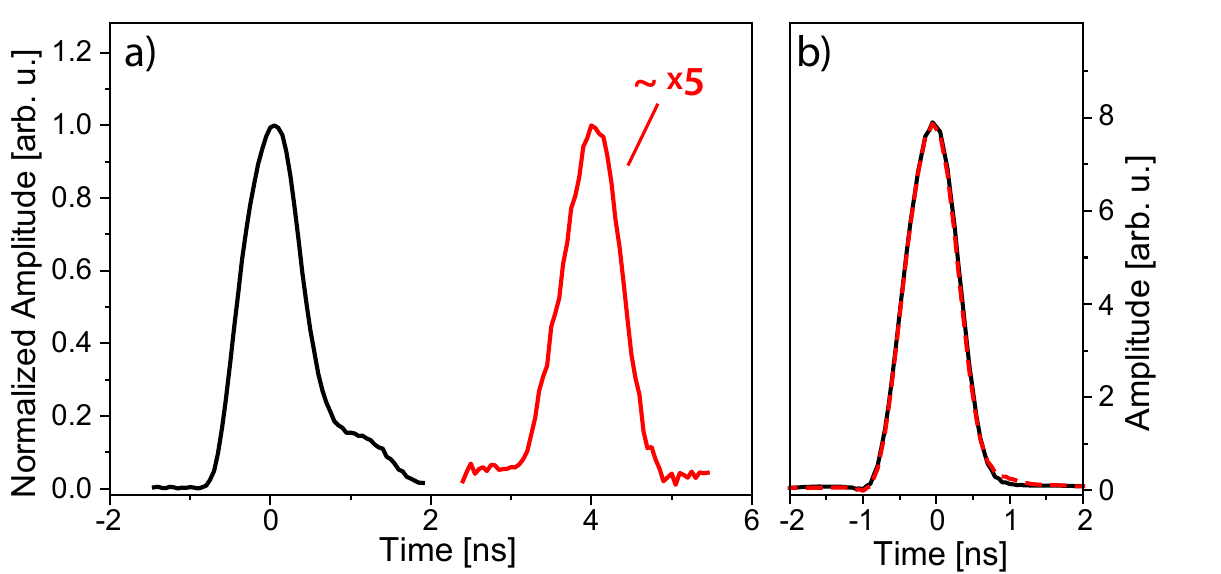}
  \caption{Non-reciprocal light storage. a) Optical data pulses (black curve) propagating in one direction are delayed by 4\,ns (red line) while b) simultaneously counter-propagating data pulses are not impacted (black line shows transmitted data, the red dotted line shows data while counter-propagating data is stored). Note, that the read-out efficiency of the pulses presented in a) is around 20\% and the data is normalized to visualise the pulse shape before and after the storage process.}
\label{stor}
\end{center}
\end{figure*}
Figure \ref{depl}\,a shows the transfer of an optical data pulse \(\omega_{\mathrm{data}}\) to an acoustic wave by a counter-propagating write pulse \(\omega_{\mathrm{w}}\) (black curve; Full-width half maximum \(\approx 1\)\,ns). The depleted optical signal is shown in red (Fig. \ref{depl}). Around 90\% of data pulse depletion could be achieved while a simultaneously counter-propagating data pulse (co-propagating to the write pulses) is not impacted by the acoustic wave generated in the depletion process (inset Fig. \ref{depl}\,a). The inset shows optical data pulses transmitted in the counter-propagating direction while there is no data transfered to the acoustic wave (black curve). The red dashed curve in the inset of Fig. \ref{depl}\,a shows the counter-propagating optical data pulses while the data in the traveling in the opposite direction is depleted and transfered to a traveling acoustic wave. Both curves perfectly overlap. In the current experimental implementation, the depletion efficiency of 90\% was mainly limited by power constraints and can in principle reach almost full conversion of the optical pulse to the acoustic wave. Figure \ref{depl}\,b shows the fast Fourier transform (FFT) of the optical input data pulses demonstrating that the bandwidth of the optical data pulses that can successfully depleted and transferred to the acoustic domain can greatly exceed the intrinsic Brillouin linewidth of 30\,MHz. The dashed lines in Fig. \ref{depl}\,b indicate the full-width half maximum of the FFT of the data pulses. \newline
This extension of the bandwidth by more than one order of magnitude beyond the intrinsic acoustic linewidth is enabled by the ultra-high Brillouin gain provided in chalcogenide waveguides which is about two orders of magnitude larger than in standard silica single-mode fiber. The broad bandwidth is one main advantage of the resonator-free waveguide-based Brillouin approach. Whereas the intrinsic linewidth of the opto-acoustic interaction is given by the phonon lifetime for a single frequency continuous wave optical pump the Brillouin response can be broadened by using a broad bandwidth optical pump where the bandwidth of the Brillouin response is then given by the bandwidth of the optical pump itself. The intrinsic Brillouin gain is in this case distributed over the bandwidth and hence the absolute maximum Brillouin gain is reduced accordingly for a given input power. \newline
After demonstrating the nonreciprocal depletion of optical data pulses we now show that we can retrieve the data back from the acoustic to the optical domain, even in the presence of counter-propagating pulses, and hence uni-directionally delay optical signals relative to the regular transit time of the waveguide. Figure \ref{stor}\,a shows experimental measurements of a 1\,ns long optical data pulse that is delayed by 4\,ns when propagating in one direction in the waveguide, while an optical data pulse simultaneously propagating in the opposite direction with the same optical frequency and optical mode is not interacting with the Brillouin storage process or the acoustic wave present in the waveguide (Fig. \ref{stor}). As a proof-of-principle demonstration, we only show a pulse delay of 4\,ns, however, we note that the delay is given by the arrival time difference between the write and the read pulses and hence can continuously be tuned. As the phonon exponentially decays as \(\mathrm{e}^{-2t/\tau_\mathrm{A}}\), with \(\tau_\mathrm{A}\) being the acoustic lifetime, the readout efficiency decreases for longer storage times. The phonon lifetime in our waveguide structure is mainly limited by the properties of the chalcogenide material. Much longer phonon lifetimes have been shown in acoustic resonators that hence could achieve longer storage times, however, at the expense of the bandwidth \cite{Fiore2011a,Fiore2013,Dong2015}. Furthermore, the shape of the pulses is challenging to be maintained in resonator-based acoustic light storage with the spatial extent of the pulses exceeding the circumference of the resonators. \newline
Conversely, in the here demonstrated delay scheme the pulse shape is maintained (Fig. \ref{stor}\,a). The input data pulse and the delayed data pulse in Fig. \ref{stor}\,a are normalized to visually emphasize that point and show the similarity in the pulse shape as it is common practice for fiber-based optical pulse delay techniques \cite{Okawachi2005,Song2005,Song2009,Preussler2009,Chin2012,Merklein2018}. The readout efficiency of this measurement was around 20\% after a delay of 4\,ns. \newline
Whereas the data pulses are stored and delayed in one direction, simultaneously counter-propagating data pulses are not delayed (Fig. \ref{stor}\,b. The black curve in Fig. \ref{stor}\,b shows an optical pulse when there is no delay applied to the data that travels in the opposite direction while the dashed red curve shows the optical pulse for the case when the counter-propagating channel is delayed via coherent transfer from optical to acoustic and back to the optical domain. Here, the data is not normalized to emphasize that there is neither a change in amplitude nor shape of the pulses. The counter-propagating data pulses do not interact with the acoustic mode present in the waveguide from the delay process of the data pulses propagating in the opposite direction nor do they distort the storage process of these pulses. Hence, we show that the non-reciprocal pulse storage scheme enables full duplex signal processing. It also shows that potential back-reflections which can occur in complex integrated circuits that consist of many discreet components is not distorting the delay process. \newline
\begin{figure}[t]
\begin{center}
  \includegraphics[width=0.7\linewidth]{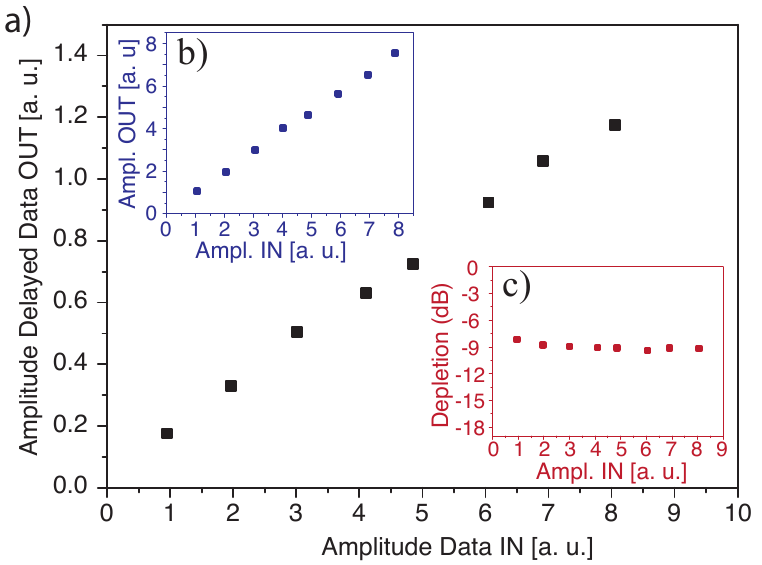}
  \caption{Linearity of non-reciprocal light storage. a) Linear amplitude response of the delayed optical data pulses. Inset b) shows the linearity of the counter-propagating data pulses while inset c) shows the depletion of the data pulses for different input amplitude levels.}
\label{lin}
\end{center}
\end{figure}
A crucial metric for non-reciprocal devices in general that is particularly important for optical signal processing applications is the linearity of the scheme. The linearity ensures that information encoded in the amplitude is maintained during the signal processing operation. Figure \ref{lin} shows that the Brillouin-based non-reciprocal delay scheme is linear over a wide range. The coupled power of the input data pulses in Fig. \ref{lin} is varied by an order of magnitude from -10\,dBm to -20\,dBm average optical power and we observe a linear relationship between in-and output amplitude. We confirm that the same is true for the counter-propagating data channel (Fig. \ref{lin}\,b). The second inset, Fig. \ref{lin}\,c, shows that the depletion of the original data pulses which are transferred to the acoustic wave is approximately constant in the measured data power range. \newline
\section{Conclusion and Outlook}
We showed a non-reciprocal delay scheme based on the coherent interaction of photons with traveling acoustic phonons. The phase-matching underlying this process ensures that only optical data pulses traveling in a distinct direction is delayed. We showed that the bandwidth of this scheme is not limited to the intrinsic linewidth of the opto-acoustic interaction, but can, in fact, be much broader approaching the GHz regime. Furthermore, we demonstrated that the scheme depends linearly on the input power and does neither convert the optical mode nor the wavelength of the signal. Hence, it opens a pathway to full duplex signal processing architectures that can greatly reduce size, weight and power requirements. The delay time is continuously tunable as it is given by the difference in the arrival time of the read out pulse with respect to the write pulse within the phonon lifetime \cite{Merklein2017}. Recently, however, it was proposed and experimentally shown that the storage time can be extended by refreshing the acoustic phonon with optical pulses \cite{Stiller2019a} overcoming said limitation.  \newline
Our demonstration of delaying an optical signal while another optical signal at the same frequency is counter-propagating shows the immunity of the here presented delay scheme to detrimental back-reflections common in complex integrated photonic circuits that are composed of a multitude of optical elements. In the context of phased array antennas and beam steering elements inducing non-reciprocal delays could enable new ways of separating transmitted and received signals.
\section{Materials and Methods}\label{methods}
\subsection{Experimental setup for non-reciprocal light storage.}
\begin{figure*}[t]
\begin{center}
  \includegraphics[width=0.8\textwidth]{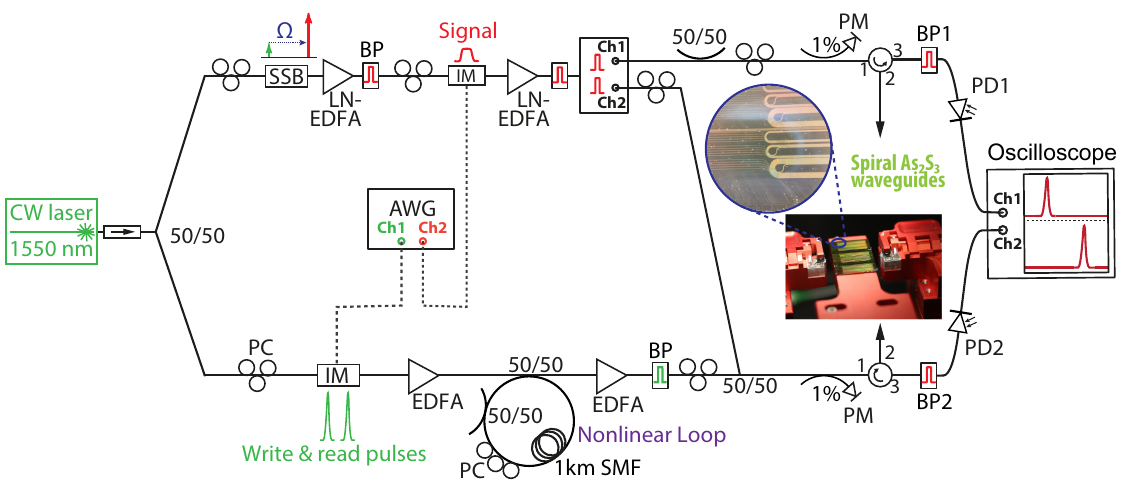}
  \caption{Experimental setup. CW laser: continuous-wave laser; 50/50: 50/50 optical fiber coupler; PC: polarisation controller; SSB: single-sideband modulator; EDFA: Erbium-doped fiber amplifier; LN-EDFA: Low-noise EDFA; IM: intensity modulator; BP: bandpass filter; AWG: arbitrary waveform generator; CH 1/\,2: Channel 1/\,2; SMF: standard single-mode fiber; PM: power meter; PD: photodetector.}
\label{detset}
\end{center}
\end{figure*}

A layout of the experimental setup is shown in Fig. \ref{detset}. 
As a laser source, we use a narrow-linewidth distributed feedback laser (TerraXion NLL) with a wavelength of around 1550\,nm. The laser signal is divided into two arms, the data and the write/\,read arm. The data signal is up-shifted in frequency by the Brillouin frequency shift \(\Omega\) via a single-sideband modulator. A single laser source is used to avoid relative drift of the data and the write/\,read arm. Two intensity modulators connected to a multi-channel arbitrary waveform generator are used to chop the continuous wave (CW) laser signals in both arms into a pulse stream. The write/\,read pulses are amplified via an erbium-doped fibre amplifier (EDFA) and afterwards pass through a nonlinear fibre loop. The nonlinear fibre loop only transmits the write/\,read pulses and suppresses any background present from the laser or amplifier between the pulses as only the pulses have a high enough intensity to induce a nonlinear phase shift in the fibre loop. Hence only the pulses are transmitted and the low-intensity background is reflected by the loop. A second EDFA after the loop boosts the signal to peak powers of several Watts. To minimise the effect of white noise bandpass filters with a bandwidth of around 0.5\,nm are implemented after every amplification step by the EDFAs. The write and read pulses are coupled into the photonic chip using lensed fibres and the average coupled on-chip power was around 7\,dBm. \newline
The data signal is split with a 50/\,50 coupler and coupled from both sides into the photonic chip with an average power of around -10\,dBm. From one side the write\,/read pulses are combined with the data pulses and coupled via the same lensed fibre into the chip. Additional 50/\,50 couplers are used in the data path to make sure the data pulses and the write/\,read pulses overlap in the middle of the waveguide. Circulators are used on both sides of the chip to route the transmitted data signal to a two-channel fast oscilloscope. Two narrowband filters with a bandwidth of 3\,GHz are used before the fast photodetectors (bandwidth 12\,GHz) to filter residual write/\,read pulses.\newline
\subsection{Storage medium.}
The storage medium for the non-reciprocal light storage is a chalcogenide rib waveguide \cite{Madden2007,Pant2011}. The chalcogenide As\(_2\)S\(_3\) thin film of around 850\,nm is deposited on a thermal oxide silicon wafer with a variation of the film thickness below 5\% \cite{Zarifi2017}. Photolithography is used to pattern the waveguide structures which are etched into the thin film using inductively-coupled plasma dry etching with a mixture of CHF\(_3\), O\(_2\), and Ar. The dimensions of the rib structure are 850\,nm by 2.2\,\(\mu\)m with a 50\% etch depth and an overall length of 22\,cm. The waveguide is arranged in a spiral to reduce the overall footprint to around 10\,mm\(^2\). The bend-radius of the spiral is around 200\,\(\mu\)m to ensure that there is no additional bending loss introduced for the fundamental optical mode as well as the acoustic mode. \newline
The chalcogenide glass As\(_2\)S\(_3\) is sandwiched between a silica substrate and a silica top-cladding to ensure guiding of the acoustic as well as the optical mode \cite{Poulton2013a}. The difference in refractive index of silica \(\mathrm{n}\approx1.4\) and the chalcogenide rib waveguides \(\mathrm{n}\approx2.4\) guarantees tight confinement of the optical mode whereas the difference in sound velocity of around 3400\,m/s prevents leakage of the acoustic mode into the substrate or cladding which enables strong overlap between the two respective modes. The silica top-cladding is deposited using sputtering. \newline 
Light is coupled in and out of the chip using lensed fibers with a roughly 2\,\(\mu\)m focus spot size. The coupling loss per facet is around 4\,dB. The polarization is adjusted so light is coupled into the fundamental TE mode of the waveguide which is the mode with the lowest loss. \newline

\textbf{Acknowledgments:} This work was supported by the Australian Research Council (ARC) through Laureate Fellowship (FL120100029), Center of Excellence CUDOS (CE110001018), ARC 2020 Discovery Project (DP200101893), ARC Linkage grant (LP170100112), and U.S. Office of Naval Research Global (ONRG) (N62909-18-1-2013). We acknowledge the support of the ANFF ACT.
\bibliography{library}

\begin{thebibliography}{46}%
\makeatletter
\providecommand \@ifxundefined [1]{%
 \@ifx{#1\undefined}
}%
\providecommand \@ifnum [1]{%
 \ifnum #1\expandafter \@firstoftwo
 \else \expandafter \@secondoftwo
 \fi
}%
\providecommand \@ifx [1]{%
 \ifx #1\expandafter \@firstoftwo
 \else \expandafter \@secondoftwo
 \fi
}%
\providecommand \natexlab [1]{#1}%
\providecommand \enquote  [1]{``#1''}%
\providecommand \bibnamefont  [1]{#1}%
\providecommand \bibfnamefont [1]{#1}%
\providecommand \citenamefont [1]{#1}%
\providecommand \href@noop [0]{\@secondoftwo}%
\providecommand \href [0]{\begingroup \@sanitize@url \@href}%
\providecommand \@href[1]{\@@startlink{#1}\@@href}%
\providecommand \@@href[1]{\endgroup#1\@@endlink}%
\providecommand \@sanitize@url [0]{\catcode `\\12\catcode `\$12\catcode
  `\&12\catcode `\#12\catcode `\^12\catcode `\_12\catcode `\%12\relax}%
\providecommand \@@startlink[1]{}%
\providecommand \@@endlink[0]{}%
\providecommand \url  [0]{\begingroup\@sanitize@url \@url }%
\providecommand \@url [1]{\endgroup\@href {#1}{\urlprefix }}%
\providecommand \urlprefix  [0]{URL }%
\providecommand \Eprint [0]{\href }%
\providecommand \doibase [0]{http://dx.doi.org/}%
\providecommand \selectlanguage [0]{\@gobble}%
\providecommand \bibinfo  [0]{\@secondoftwo}%
\providecommand \bibfield  [0]{\@secondoftwo}%
\providecommand \translation [1]{[#1]}%
\providecommand \BibitemOpen [0]{}%
\providecommand \bibitemStop [0]{}%
\providecommand \bibitemNoStop [0]{.\EOS\space}%
\providecommand \EOS [0]{\spacefactor3000\relax}%
\providecommand \BibitemShut  [1]{\csname bibitem#1\endcsname}%
\let\auto@bib@innerbib\@empty
\bibitem [{\citenamefont {Jalas}\ \emph {et~al.}(2013)\citenamefont {Jalas},
  \citenamefont {Petrov}, \citenamefont {Eich}, \citenamefont {Freude},
  \citenamefont {Fan}, \citenamefont {Yu}, \citenamefont {Baets}, \citenamefont
  {Popovi{\'{c}}}, \citenamefont {Melloni}, \citenamefont {Joannopoulos},
  \citenamefont {Vanwolleghem}, \citenamefont {Doerr},\ and\ \citenamefont
  {Renner}}]{Jalas2013}%
  \BibitemOpen
  \bibfield  {author} {\bibinfo {author} {\bibfnamefont {Dirk}\ \bibnamefont
  {Jalas}}, \bibinfo {author} {\bibfnamefont {Alexander}\ \bibnamefont
  {Petrov}}, \bibinfo {author} {\bibfnamefont {Manfred}\ \bibnamefont {Eich}},
  \bibinfo {author} {\bibfnamefont {Wolfgang}\ \bibnamefont {Freude}}, \bibinfo
  {author} {\bibfnamefont {Shanhui}\ \bibnamefont {Fan}}, \bibinfo {author}
  {\bibfnamefont {Zongfu}\ \bibnamefont {Yu}}, \bibinfo {author} {\bibfnamefont
  {Roel}\ \bibnamefont {Baets}}, \bibinfo {author} {\bibfnamefont
  {Milo{\v{s}}}\ \bibnamefont {Popovi{\'{c}}}}, \bibinfo {author}
  {\bibfnamefont {Andrea}\ \bibnamefont {Melloni}}, \bibinfo {author}
  {\bibfnamefont {John~D.}\ \bibnamefont {Joannopoulos}}, \bibinfo {author}
  {\bibfnamefont {Mathias}\ \bibnamefont {Vanwolleghem}}, \bibinfo {author}
  {\bibfnamefont {Christopher~R.}\ \bibnamefont {Doerr}}, \ and\ \bibinfo
  {author} {\bibfnamefont {Hagen}\ \bibnamefont {Renner}},\ }\bibfield  {title}
  {\enquote {\bibinfo {title} {{What is — and what is not — an optical
  isolator}},}\ }\href {\doibase 10.1038/nphoton.2013.185} {\bibfield
  {journal} {\bibinfo  {journal} {Nature Photonics}\ }\textbf {\bibinfo
  {volume} {7}},\ \bibinfo {pages} {579--582} (\bibinfo {year}
  {2013})}\BibitemShut {NoStop}%
\bibitem [{\citenamefont {Sounas}\ and\ \citenamefont
  {Al{\`{u}}}(2017)}]{2017}%
  \BibitemOpen
  \bibfield  {author} {\bibinfo {author} {\bibfnamefont {Dimitrios~L.}\
  \bibnamefont {Sounas}}\ and\ \bibinfo {author} {\bibfnamefont {Andrea}\
  \bibnamefont {Al{\`{u}}}},\ }\bibfield  {title} {\enquote {\bibinfo {title}
  {{Non-reciprocal photonics based on time modulation}},}\ }\href {\doibase
  10.1038/s41566-017-0051-x} {\bibfield  {journal} {\bibinfo  {journal} {Nature
  Photonics}\ }\textbf {\bibinfo {volume} {11}},\ \bibinfo {pages} {774--783}
  (\bibinfo {year} {2017})}\BibitemShut {NoStop}%
\bibitem [{\citenamefont {Caloz}\ \emph {et~al.}(2018)\citenamefont {Caloz},
  \citenamefont {Al{\`{u}}}, \citenamefont {Tretyakov}, \citenamefont {Sounas},
  \citenamefont {Achouri},\ and\ \citenamefont {Deck-L{\'{e}}ger}}]{Caloz2018}%
  \BibitemOpen
  \bibfield  {author} {\bibinfo {author} {\bibfnamefont {Christophe}\
  \bibnamefont {Caloz}}, \bibinfo {author} {\bibfnamefont {Andrea}\
  \bibnamefont {Al{\`{u}}}}, \bibinfo {author} {\bibfnamefont {Sergei}\
  \bibnamefont {Tretyakov}}, \bibinfo {author} {\bibfnamefont {Dimitrios}\
  \bibnamefont {Sounas}}, \bibinfo {author} {\bibfnamefont {Karim}\
  \bibnamefont {Achouri}}, \ and\ \bibinfo {author} {\bibfnamefont
  {Zo{\'{e}}~Lise}\ \bibnamefont {Deck-L{\'{e}}ger}},\ }\bibfield  {title}
  {\enquote {\bibinfo {title} {{Electromagnetic Nonreciprocity}},}\ }\href
  {\doibase 10.1103/PhysRevApplied.10.047001} {\bibfield  {journal} {\bibinfo
  {journal} {Physical Review Applied}\ }\textbf {\bibinfo {volume} {10}},\
  \bibinfo {pages} {1} (\bibinfo {year} {2018})}\BibitemShut {NoStop}%
\bibitem [{\citenamefont {Shoji}\ \emph {et~al.}(2008)\citenamefont {Shoji},
  \citenamefont {Mizumoto}, \citenamefont {Yokoi}, \citenamefont {Hsieh},\ and\
  \citenamefont {Osgood}}]{Shoji2008}%
  \BibitemOpen
  \bibfield  {author} {\bibinfo {author} {\bibfnamefont {Yuya}\ \bibnamefont
  {Shoji}}, \bibinfo {author} {\bibfnamefont {Tetsuya}\ \bibnamefont
  {Mizumoto}}, \bibinfo {author} {\bibfnamefont {Hideki}\ \bibnamefont
  {Yokoi}}, \bibinfo {author} {\bibfnamefont {I.~Wei}\ \bibnamefont {Hsieh}}, \
  and\ \bibinfo {author} {\bibfnamefont {Richard~M.}\ \bibnamefont {Osgood}},\
  }\bibfield  {title} {\enquote {\bibinfo {title} {{Magneto-optical isolator
  with silicon waveguides fabricated by direct bonding}},}\ }\href {\doibase
  10.1063/1.2884855} {\bibfield  {journal} {\bibinfo  {journal} {Applied
  Physics Letters}\ }\textbf {\bibinfo {volume} {92}},\ \bibinfo {pages} {2--5}
  (\bibinfo {year} {2008})}\BibitemShut {NoStop}%
\bibitem [{\citenamefont {Bi}\ \emph {et~al.}(2011)\citenamefont {Bi},
  \citenamefont {Hu}, \citenamefont {Jiang}, \citenamefont {Kim}, \citenamefont
  {Dionne}, \citenamefont {Kimerling},\ and\ \citenamefont {Ross}}]{Bi2011}%
  \BibitemOpen
  \bibfield  {author} {\bibinfo {author} {\bibfnamefont {Lei}\ \bibnamefont
  {Bi}}, \bibinfo {author} {\bibfnamefont {Juejun}\ \bibnamefont {Hu}},
  \bibinfo {author} {\bibfnamefont {Peng}\ \bibnamefont {Jiang}}, \bibinfo
  {author} {\bibfnamefont {Dong~Hun}\ \bibnamefont {Kim}}, \bibinfo {author}
  {\bibfnamefont {Gerald~F.}\ \bibnamefont {Dionne}}, \bibinfo {author}
  {\bibfnamefont {Lionel~C.}\ \bibnamefont {Kimerling}}, \ and\ \bibinfo
  {author} {\bibfnamefont {C.~a.}\ \bibnamefont {Ross}},\ }\bibfield  {title}
  {\enquote {\bibinfo {title} {{On-chip optical isolation in monolithically
  integrated non-reciprocal optical resonators}},}\ }\href {\doibase
  10.1038/nphoton.2011.270} {\bibfield  {journal} {\bibinfo  {journal} {Nature
  Photonics}\ }\textbf {\bibinfo {volume} {5}},\ \bibinfo {pages} {758--762}
  (\bibinfo {year} {2011})}\BibitemShut {NoStop}%
\bibitem [{\citenamefont {Huang}\ \emph {et~al.}(2017)\citenamefont {Huang},
  \citenamefont {Pintus}, \citenamefont {Zhang}, \citenamefont {Morton},
  \citenamefont {Shoji}, \citenamefont {Mizumoto},\ and\ \citenamefont
  {Bowers}}]{Huang2017a}%
  \BibitemOpen
  \bibfield  {author} {\bibinfo {author} {\bibfnamefont {Duanni}\ \bibnamefont
  {Huang}}, \bibinfo {author} {\bibfnamefont {Paolo}\ \bibnamefont {Pintus}},
  \bibinfo {author} {\bibfnamefont {Chong}\ \bibnamefont {Zhang}}, \bibinfo
  {author} {\bibfnamefont {Paul}\ \bibnamefont {Morton}}, \bibinfo {author}
  {\bibfnamefont {Yuya}\ \bibnamefont {Shoji}}, \bibinfo {author}
  {\bibfnamefont {Tetsuya}\ \bibnamefont {Mizumoto}}, \ and\ \bibinfo {author}
  {\bibfnamefont {John~E.}\ \bibnamefont {Bowers}},\ }\bibfield  {title}
  {\enquote {\bibinfo {title} {{Dynamically reconfigurable integrated optical
  circulators}},}\ }\href {\doibase 10.1364/OPTICA.4.000023} {\bibfield
  {journal} {\bibinfo  {journal} {Optica}\ }\textbf {\bibinfo {volume} {4}},\
  \bibinfo {pages} {23} (\bibinfo {year} {2017})}\BibitemShut {NoStop}%
\bibitem [{\citenamefont {Yu}\ and\ \citenamefont {Fan}(2009)}]{Yu2009}%
  \BibitemOpen
  \bibfield  {author} {\bibinfo {author} {\bibfnamefont {Zongfu}\ \bibnamefont
  {Yu}}\ and\ \bibinfo {author} {\bibfnamefont {Shanhui}\ \bibnamefont {Fan}},\
  }\bibfield  {title} {\enquote {\bibinfo {title} {{Complete optical isolation
  created by indirect interband photonic transitions}},}\ }\href {\doibase
  10.1038/nphoton.2008.273} {\bibfield  {journal} {\bibinfo  {journal} {Nature
  Photonics}\ }\textbf {\bibinfo {volume} {3}},\ \bibinfo {pages} {91--94}
  (\bibinfo {year} {2009})}\BibitemShut {NoStop}%
\bibitem [{\citenamefont {Lira}\ \emph {et~al.}(2012)\citenamefont {Lira},
  \citenamefont {Yu}, \citenamefont {Fan},\ and\ \citenamefont
  {Lipson}}]{Lira2012}%
  \BibitemOpen
  \bibfield  {author} {\bibinfo {author} {\bibfnamefont {Hugo}\ \bibnamefont
  {Lira}}, \bibinfo {author} {\bibfnamefont {Zongfu}\ \bibnamefont {Yu}},
  \bibinfo {author} {\bibfnamefont {Shanhui}\ \bibnamefont {Fan}}, \ and\
  \bibinfo {author} {\bibfnamefont {Michal}\ \bibnamefont {Lipson}},\
  }\bibfield  {title} {\enquote {\bibinfo {title} {{Electrically driven
  nonreciprocity induced by interband photonic transition on a silicon
  chip}},}\ }\href {\doibase 10.1103/PhysRevLett.109.033901} {\bibfield
  {journal} {\bibinfo  {journal} {Physical Review Letters}\ }\textbf {\bibinfo
  {volume} {109}},\ \bibinfo {pages} {1--5} (\bibinfo {year}
  {2012})}\BibitemShut {NoStop}%
\bibitem [{\citenamefont {Kang}\ \emph {et~al.}(2011)\citenamefont {Kang},
  \citenamefont {Butsch},\ and\ \citenamefont {Russell}}]{Kang2011}%
  \BibitemOpen
  \bibfield  {author} {\bibinfo {author} {\bibfnamefont {M~S}\ \bibnamefont
  {Kang}}, \bibinfo {author} {\bibfnamefont {A}~\bibnamefont {Butsch}}, \ and\
  \bibinfo {author} {\bibfnamefont {P~St~J}\ \bibnamefont {Russell}},\
  }\bibfield  {title} {\enquote {\bibinfo {title} {{Reconfigurable light-driven
  opto-acoustic isolators in photonic crystal fibre}},}\ }\href {\doibase
  10.1038/nphoton.2011.180} {\bibfield  {journal} {\bibinfo  {journal} {Nature
  Photonics}\ }\textbf {\bibinfo {volume} {5}},\ \bibinfo {pages} {549--553}
  (\bibinfo {year} {2011})}\BibitemShut {NoStop}%
\bibitem [{\citenamefont {Sohn}\ \emph {et~al.}(2018)\citenamefont {Sohn},
  \citenamefont {Kim},\ and\ \citenamefont {Bahl}}]{Sohn2018}%
  \BibitemOpen
  \bibfield  {author} {\bibinfo {author} {\bibfnamefont {Donggyu~B}\
  \bibnamefont {Sohn}}, \bibinfo {author} {\bibfnamefont {Seunghwi}\
  \bibnamefont {Kim}}, \ and\ \bibinfo {author} {\bibfnamefont {Gaurav}\
  \bibnamefont {Bahl}},\ }\bibfield  {title} {\enquote {\bibinfo {title}
  {{Time-reversal symmetry breaking with acoustic pumping of nanophotonic
  circuits}},}\ }\href {\doibase 10.1038/s41566-017-0075-2} {\bibfield
  {journal} {\bibinfo  {journal} {Nature Photonics}\ }\textbf {\bibinfo
  {volume} {12}},\ \bibinfo {pages} {91--97} (\bibinfo {year}
  {2018})}\BibitemShut {NoStop}%
\bibitem [{\citenamefont {Verhagen}\ and\ \citenamefont
  {Al{\`{u}}}(2017)}]{Verhagen2017}%
  \BibitemOpen
  \bibfield  {author} {\bibinfo {author} {\bibfnamefont {Ewold}\ \bibnamefont
  {Verhagen}}\ and\ \bibinfo {author} {\bibfnamefont {Andrea}\ \bibnamefont
  {Al{\`{u}}}},\ }\bibfield  {title} {\enquote {\bibinfo {title}
  {{Optomechanical nonreciprocity}},}\ }\href {\doibase 10.1038/nphys4283}
  {\bibfield  {journal} {\bibinfo  {journal} {Nature Physics}\ }\textbf
  {\bibinfo {volume} {13}},\ \bibinfo {pages} {922--924} (\bibinfo {year}
  {2017})}\BibitemShut {NoStop}%
\bibitem [{\citenamefont {Miri}\ \emph {et~al.}(2017)\citenamefont {Miri},
  \citenamefont {Ruesink}, \citenamefont {Verhagen},\ and\ \citenamefont
  {Al{\`{u}}}}]{Miri2017}%
  \BibitemOpen
  \bibfield  {author} {\bibinfo {author} {\bibfnamefont {Mohammad~Ali}\
  \bibnamefont {Miri}}, \bibinfo {author} {\bibfnamefont {Freek}\ \bibnamefont
  {Ruesink}}, \bibinfo {author} {\bibfnamefont {Ewold}\ \bibnamefont
  {Verhagen}}, \ and\ \bibinfo {author} {\bibfnamefont {Andrea}\ \bibnamefont
  {Al{\`{u}}}},\ }\bibfield  {title} {\enquote {\bibinfo {title} {{Optical
  Nonreciprocity Based on Optomechanical Coupling}},}\ }\href {\doibase
  10.1103/PhysRevApplied.7.064014} {\bibfield  {journal} {\bibinfo  {journal}
  {Physical Review Applied}\ }\textbf {\bibinfo {volume} {7}},\ \bibinfo
  {pages} {1--20} (\bibinfo {year} {2017})}\BibitemShut {NoStop}%
\bibitem [{\citenamefont {Manipatruni}\ \emph {et~al.}(2009)\citenamefont
  {Manipatruni}, \citenamefont {Robinson},\ and\ \citenamefont
  {Lipson}}]{Achar2008}%
  \BibitemOpen
  \bibfield  {author} {\bibinfo {author} {\bibfnamefont {Sasikanth}\
  \bibnamefont {Manipatruni}}, \bibinfo {author} {\bibfnamefont {Jacob~T.}\
  \bibnamefont {Robinson}}, \ and\ \bibinfo {author} {\bibfnamefont {Michal}\
  \bibnamefont {Lipson}},\ }\bibfield  {title} {\enquote {\bibinfo {title}
  {{Optical Nonreciprocity in Optomechanical Structures}},}\ }\href {\doibase
  10.1103/PhysRevLett.102.213903} {\bibfield  {journal} {\bibinfo  {journal}
  {Physical Review Letters}\ }\textbf {\bibinfo {volume} {102}},\ \bibinfo
  {pages} {213903} (\bibinfo {year} {2009})}\BibitemShut {NoStop}%
\bibitem [{\citenamefont {Hafezi}\ and\ \citenamefont
  {Rabl}(2012)}]{Verhagen2012a}%
  \BibitemOpen
  \bibfield  {author} {\bibinfo {author} {\bibfnamefont {Mohammad}\
  \bibnamefont {Hafezi}}\ and\ \bibinfo {author} {\bibfnamefont {Peter}\
  \bibnamefont {Rabl}},\ }\bibfield  {title} {\enquote {\bibinfo {title}
  {{Optomechanically induced non-reciprocity in microring resonators}},}\
  }\href {\doibase 10.1364/OE.20.007672} {\bibfield  {journal} {\bibinfo
  {journal} {Optics Express}\ }\textbf {\bibinfo {volume} {20}},\ \bibinfo
  {pages} {7672} (\bibinfo {year} {2012})}\BibitemShut {NoStop}%
\bibitem [{\citenamefont {Xu}\ and\ \citenamefont {Li}(2015)}]{Xu2015b}%
  \BibitemOpen
  \bibfield  {author} {\bibinfo {author} {\bibfnamefont {Xun~Wei}\ \bibnamefont
  {Xu}}\ and\ \bibinfo {author} {\bibfnamefont {Yong}\ \bibnamefont {Li}},\
  }\bibfield  {title} {\enquote {\bibinfo {title} {{Optical nonreciprocity and
  optomechanical circulator in three-mode optomechanical systems}},}\ }\href
  {\doibase 10.1103/PhysRevA.91.053854} {\bibfield  {journal} {\bibinfo
  {journal} {Physical Review A - Atomic, Molecular, and Optical Physics}\
  }\textbf {\bibinfo {volume} {91}},\ \bibinfo {pages} {1--8} (\bibinfo {year}
  {2015})}\BibitemShut {NoStop}%
\bibitem [{\citenamefont {Ruesink}\ \emph {et~al.}(2016)\citenamefont
  {Ruesink}, \citenamefont {Miri}, \citenamefont {Al{\`{u}}},\ and\
  \citenamefont {Verhagen}}]{Ruesink2016}%
  \BibitemOpen
  \bibfield  {author} {\bibinfo {author} {\bibfnamefont {Freek}\ \bibnamefont
  {Ruesink}}, \bibinfo {author} {\bibfnamefont {Mohammad-Ali}\ \bibnamefont
  {Miri}}, \bibinfo {author} {\bibfnamefont {Andrea}\ \bibnamefont
  {Al{\`{u}}}}, \ and\ \bibinfo {author} {\bibfnamefont {Ewold}\ \bibnamefont
  {Verhagen}},\ }\bibfield  {title} {\enquote {\bibinfo {title}
  {{Nonreciprocity and magnetic-free isolation based on optomechanical
  interactions}},}\ }\href {\doibase 1DOI: 0.1038/ncomms13662} {\bibfield
  {journal} {\bibinfo  {journal} {Nature Communications}\ }\textbf {\bibinfo
  {volume} {7}},\ \bibinfo {pages} {13662} (\bibinfo {year}
  {2016})}\BibitemShut {NoStop}%
\bibitem [{\citenamefont {Shen}\ \emph {et~al.}(2016)\citenamefont {Shen},
  \citenamefont {Zhang}, \citenamefont {Chen}, \citenamefont {Zou},
  \citenamefont {Xiao}, \citenamefont {Zou}, \citenamefont {Sun}, \citenamefont
  {Guo},\ and\ \citenamefont {Dong}}]{Shen2016}%
  \BibitemOpen
  \bibfield  {author} {\bibinfo {author} {\bibfnamefont {Zhen}\ \bibnamefont
  {Shen}}, \bibinfo {author} {\bibfnamefont {Yan-Lei}\ \bibnamefont {Zhang}},
  \bibinfo {author} {\bibfnamefont {Yuan}\ \bibnamefont {Chen}}, \bibinfo
  {author} {\bibfnamefont {Chang-Ling}\ \bibnamefont {Zou}}, \bibinfo {author}
  {\bibfnamefont {Yun-Feng}\ \bibnamefont {Xiao}}, \bibinfo {author}
  {\bibfnamefont {Xu-Bo}\ \bibnamefont {Zou}}, \bibinfo {author} {\bibfnamefont
  {Fang-Wen}\ \bibnamefont {Sun}}, \bibinfo {author} {\bibfnamefont
  {Guang-Can}\ \bibnamefont {Guo}}, \ and\ \bibinfo {author} {\bibfnamefont
  {Chun-Hua}\ \bibnamefont {Dong}},\ }\bibfield  {title} {\enquote {\bibinfo
  {title} {{Experimental realization of optomechanically induced
  non-reciprocity}},}\ }\href {\doibase 10.1038/nphoton.2016.161} {\bibfield
  {journal} {\bibinfo  {journal} {Nature Photonics}\ }\textbf {\bibinfo
  {volume} {10}},\ \bibinfo {pages} {657--661} (\bibinfo {year}
  {2016})}\BibitemShut {NoStop}%
\bibitem [{\citenamefont {Fang}\ \emph {et~al.}(2017)\citenamefont {Fang},
  \citenamefont {Luo}, \citenamefont {Metelmann}, \citenamefont {Matheny},
  \citenamefont {Marquardt}, \citenamefont {Clerk},\ and\ \citenamefont
  {Painter}}]{Fang2016b}%
  \BibitemOpen
  \bibfield  {author} {\bibinfo {author} {\bibfnamefont {Kejie}\ \bibnamefont
  {Fang}}, \bibinfo {author} {\bibfnamefont {Jie}\ \bibnamefont {Luo}},
  \bibinfo {author} {\bibfnamefont {Anja}\ \bibnamefont {Metelmann}}, \bibinfo
  {author} {\bibfnamefont {Matthew~H.}\ \bibnamefont {Matheny}}, \bibinfo
  {author} {\bibfnamefont {Florian}\ \bibnamefont {Marquardt}}, \bibinfo
  {author} {\bibfnamefont {Aashish~A.}\ \bibnamefont {Clerk}}, \ and\ \bibinfo
  {author} {\bibfnamefont {Oskar}\ \bibnamefont {Painter}},\ }\bibfield
  {title} {\enquote {\bibinfo {title} {{Generalized non-reciprocity in an
  optomechanical circuit via synthetic magnetism and reservoir engineering}},}\
  }\href {\doibase 10.1038/nphys4009} {\bibfield  {journal} {\bibinfo
  {journal} {Nature Physics}\ ,\ \bibinfo {pages} {1--18}} (\bibinfo {year}
  {2017})}\BibitemShut {NoStop}%
\bibitem [{\citenamefont {Shen}\ \emph {et~al.}(2018)\citenamefont {Shen},
  \citenamefont {Zhang}, \citenamefont {Chen}, \citenamefont {Sun},
  \citenamefont {Zou}, \citenamefont {Guo}, \citenamefont {Zou},\ and\
  \citenamefont {Dong}}]{Shen2018}%
  \BibitemOpen
  \bibfield  {author} {\bibinfo {author} {\bibfnamefont {Zhen}\ \bibnamefont
  {Shen}}, \bibinfo {author} {\bibfnamefont {Yan-Lei}\ \bibnamefont {Zhang}},
  \bibinfo {author} {\bibfnamefont {Yuan}\ \bibnamefont {Chen}}, \bibinfo
  {author} {\bibfnamefont {Fang-Wen}\ \bibnamefont {Sun}}, \bibinfo {author}
  {\bibfnamefont {Xu-Bo}\ \bibnamefont {Zou}}, \bibinfo {author} {\bibfnamefont
  {Guang-Can}\ \bibnamefont {Guo}}, \bibinfo {author} {\bibfnamefont
  {Chang-Ling}\ \bibnamefont {Zou}}, \ and\ \bibinfo {author} {\bibfnamefont
  {Chun-Hua}\ \bibnamefont {Dong}},\ }\bibfield  {title} {\enquote {\bibinfo
  {title} {{Reconfigurable optomechanical circulator and directional
  amplifier}},}\ }\href {\doibase 10.1038/s41467-018-04187-8} {\bibfield
  {journal} {\bibinfo  {journal} {Nature Communications}\ }\textbf {\bibinfo
  {volume} {9}},\ \bibinfo {pages} {1797} (\bibinfo {year} {2018})}\BibitemShut
  {NoStop}%
\bibitem [{\citenamefont {Estep}\ \emph {et~al.}(2016)\citenamefont {Estep},
  \citenamefont {Sounas},\ and\ \citenamefont {Al{\`{u}}}}]{Estep2016}%
  \BibitemOpen
  \bibfield  {author} {\bibinfo {author} {\bibfnamefont {Nicholas~Aaron}\
  \bibnamefont {Estep}}, \bibinfo {author} {\bibfnamefont {Dimitrios~L.}\
  \bibnamefont {Sounas}}, \ and\ \bibinfo {author} {\bibfnamefont {Andrea}\
  \bibnamefont {Al{\`{u}}}},\ }\bibfield  {title} {\enquote {\bibinfo {title}
  {{Magnetless microwave circulators based on spatiotemporally modulated rings
  of coupled resonators}},}\ }\href {\doibase 10.1109/TMTT.2015.2511737}
  {\bibfield  {journal} {\bibinfo  {journal} {IEEE Transactions on Microwave
  Theory and Techniques}\ }\textbf {\bibinfo {volume} {64}},\ \bibinfo {pages}
  {502--518} (\bibinfo {year} {2016})}\BibitemShut {NoStop}%
\bibitem [{\citenamefont {Peterson}\ \emph {et~al.}(2017)\citenamefont
  {Peterson}, \citenamefont {Lecocq}, \citenamefont {Cicak}, \citenamefont
  {Simmonds}, \citenamefont {Aumentado},\ and\ \citenamefont
  {Teufel}}]{Peterson2017b}%
  \BibitemOpen
  \bibfield  {author} {\bibinfo {author} {\bibfnamefont {G.~A.}\ \bibnamefont
  {Peterson}}, \bibinfo {author} {\bibfnamefont {F}~\bibnamefont {Lecocq}},
  \bibinfo {author} {\bibfnamefont {K}~\bibnamefont {Cicak}}, \bibinfo {author}
  {\bibfnamefont {R.~W.}\ \bibnamefont {Simmonds}}, \bibinfo {author}
  {\bibfnamefont {J}~\bibnamefont {Aumentado}}, \ and\ \bibinfo {author}
  {\bibfnamefont {J.~D.}\ \bibnamefont {Teufel}},\ }\bibfield  {title}
  {\enquote {\bibinfo {title} {{Demonstration of Efficient Nonreciprocity in a
  Microwave Optomechanical Circuit}},}\ }\href {\doibase
  10.1103/PhysRevX.7.031001} {\bibfield  {journal} {\bibinfo  {journal}
  {Physical Review X}\ }\textbf {\bibinfo {volume} {7}},\ \bibinfo {pages}
  {031001} (\bibinfo {year} {2017})}\BibitemShut {NoStop}%
\bibitem [{\citenamefont {Barzanjeh}\ \emph {et~al.}(2017)\citenamefont
  {Barzanjeh}, \citenamefont {Wulf}, \citenamefont {Peruzzo}, \citenamefont
  {Kalaee}, \citenamefont {Dieterle}, \citenamefont {Painter},\ and\
  \citenamefont {Fink}}]{Barzanjeh2017}%
  \BibitemOpen
  \bibfield  {author} {\bibinfo {author} {\bibfnamefont {S}~\bibnamefont
  {Barzanjeh}}, \bibinfo {author} {\bibfnamefont {M}~\bibnamefont {Wulf}},
  \bibinfo {author} {\bibfnamefont {M}~\bibnamefont {Peruzzo}}, \bibinfo
  {author} {\bibfnamefont {M}~\bibnamefont {Kalaee}}, \bibinfo {author}
  {\bibfnamefont {P~B}\ \bibnamefont {Dieterle}}, \bibinfo {author}
  {\bibfnamefont {O}~\bibnamefont {Painter}}, \ and\ \bibinfo {author}
  {\bibfnamefont {J~M}\ \bibnamefont {Fink}},\ }\bibfield  {title} {\enquote
  {\bibinfo {title} {{Mechanical on-chip microwave circulator}},}\ }\href
  {\doibase 10.1038/s41467-017-01304-x} {\bibfield  {journal} {\bibinfo
  {journal} {Nature Communications}\ }\textbf {\bibinfo {volume} {8}},\
  \bibinfo {pages} {953} (\bibinfo {year} {2017})}\BibitemShut {NoStop}%
\bibitem [{\citenamefont {Bernier}\ \emph {et~al.}(2017)\citenamefont
  {Bernier}, \citenamefont {T{\'{o}}th}, \citenamefont {Koottandavida},
  \citenamefont {Ioannou}, \citenamefont {Malz}, \citenamefont {Nunnenkamp},
  \citenamefont {Feofanov},\ and\ \citenamefont {Kippenberg}}]{Bernier2017}%
  \BibitemOpen
  \bibfield  {author} {\bibinfo {author} {\bibfnamefont {N.~R.}\ \bibnamefont
  {Bernier}}, \bibinfo {author} {\bibfnamefont {L.~D.}\ \bibnamefont
  {T{\'{o}}th}}, \bibinfo {author} {\bibfnamefont {A.}~\bibnamefont
  {Koottandavida}}, \bibinfo {author} {\bibfnamefont {M.~A.}\ \bibnamefont
  {Ioannou}}, \bibinfo {author} {\bibfnamefont {D.}~\bibnamefont {Malz}},
  \bibinfo {author} {\bibfnamefont {A.}~\bibnamefont {Nunnenkamp}}, \bibinfo
  {author} {\bibfnamefont {A.~K.}\ \bibnamefont {Feofanov}}, \ and\ \bibinfo
  {author} {\bibfnamefont {T.~J.}\ \bibnamefont {Kippenberg}},\ }\bibfield
  {title} {\enquote {\bibinfo {title} {{Nonreciprocal reconfigurable microwave
  optomechanical circuit}},}\ }\href {\doibase 10.1038/s41467-017-00447-1}
  {\bibfield  {journal} {\bibinfo  {journal} {Nature Communications}\ }\textbf
  {\bibinfo {volume} {8}} (\bibinfo {year} {2017}),\
  10.1038/s41467-017-00447-1}\BibitemShut {NoStop}%
\bibitem [{\citenamefont {Huang}\ and\ \citenamefont {Fan}(2011)}]{Huang2011}%
  \BibitemOpen
  \bibfield  {author} {\bibinfo {author} {\bibfnamefont {Xinpeng}\ \bibnamefont
  {Huang}}\ and\ \bibinfo {author} {\bibfnamefont {Shanhui}\ \bibnamefont
  {Fan}},\ }\bibfield  {title} {\enquote {\bibinfo {title} {{Complete
  all-optical silica fiber isolator via stimulated Brillouin scattering}},}\
  }\href {\doibase 10.1109/JLT.2011.2158886} {\bibfield  {journal} {\bibinfo
  {journal} {Journal of Lightwave Technology}\ }\textbf {\bibinfo {volume}
  {29}},\ \bibinfo {pages} {2267--2275} (\bibinfo {year} {2011})}\BibitemShut
  {NoStop}%
\bibitem [{\citenamefont {Eggleton}\ \emph {et~al.}(2019)\citenamefont
  {Eggleton}, \citenamefont {Poulton}, \citenamefont {Rakich}, \citenamefont
  {Steel},\ and\ \citenamefont {Bahl}}]{Eggleton2019}%
  \BibitemOpen
  \bibfield  {author} {\bibinfo {author} {\bibfnamefont {Benjamin~J.}\
  \bibnamefont {Eggleton}}, \bibinfo {author} {\bibfnamefont {Christopher~G.}\
  \bibnamefont {Poulton}}, \bibinfo {author} {\bibfnamefont {Peter~T.}\
  \bibnamefont {Rakich}}, \bibinfo {author} {\bibfnamefont {Michael.~J.}\
  \bibnamefont {Steel}}, \ and\ \bibinfo {author} {\bibfnamefont {Gaurav}\
  \bibnamefont {Bahl}},\ }\bibfield  {title} {\enquote {\bibinfo {title}
  {{Brillouin integrated photonics}},}\ }\href {\doibase
  10.1038/s41566-019-0498-z} {\bibfield  {journal} {\bibinfo  {journal} {Nature
  Photonics}\ }\textbf {\bibinfo {volume} {13}} (\bibinfo {year} {2019}),\
  10.1038/s41566-019-0498-z}\BibitemShut {NoStop}%
\bibitem [{\citenamefont {Poulton}\ \emph {et~al.}(2012)\citenamefont
  {Poulton}, \citenamefont {Pant}, \citenamefont {Byrnes}, \citenamefont {Fan},
  \citenamefont {Steel},\ and\ \citenamefont {Eggleton}}]{Poulton2012c}%
  \BibitemOpen
  \bibfield  {author} {\bibinfo {author} {\bibfnamefont {Christopher~G.}\
  \bibnamefont {Poulton}}, \bibinfo {author} {\bibfnamefont {Ravi}\
  \bibnamefont {Pant}}, \bibinfo {author} {\bibfnamefont {Adam}\ \bibnamefont
  {Byrnes}}, \bibinfo {author} {\bibfnamefont {Shanhui}\ \bibnamefont {Fan}},
  \bibinfo {author} {\bibfnamefont {M.~J.}\ \bibnamefont {Steel}}, \ and\
  \bibinfo {author} {\bibfnamefont {Benjamin~J.}\ \bibnamefont {Eggleton}},\
  }\bibfield  {title} {\enquote {\bibinfo {title} {{Design for broadband
  on-chip isolator using stimulated Brillouin scattering in
  dispersion-engineered chalcogenide waveguides}},}\ }\href {\doibase
  10.1364/OE.20.021235} {\bibfield  {journal} {\bibinfo  {journal} {Optics
  Express}\ }\textbf {\bibinfo {volume} {20}},\ \bibinfo {pages} {21235}
  (\bibinfo {year} {2012})}\BibitemShut {NoStop}%
\bibitem [{\citenamefont {Kittlaus}\ \emph {et~al.}(2018)\citenamefont
  {Kittlaus}, \citenamefont {Otterstrom}, \citenamefont {Kharel}, \citenamefont
  {Gertler},\ and\ \citenamefont {Rakich}}]{Kittlaus2018}%
  \BibitemOpen
  \bibfield  {author} {\bibinfo {author} {\bibfnamefont {Eric~A.}\ \bibnamefont
  {Kittlaus}}, \bibinfo {author} {\bibfnamefont {Nils~T.}\ \bibnamefont
  {Otterstrom}}, \bibinfo {author} {\bibfnamefont {Prashanta}\ \bibnamefont
  {Kharel}}, \bibinfo {author} {\bibfnamefont {Shai}\ \bibnamefont {Gertler}},
  \ and\ \bibinfo {author} {\bibfnamefont {Peter~T.}\ \bibnamefont {Rakich}},\
  }\bibfield  {title} {\enquote {\bibinfo {title} {{Non-reciprocal interband
  Brillouin modulation}},}\ }\href {\doibase 10.1038/s41566-018-0254-9}
  {\bibfield  {journal} {\bibinfo  {journal} {Nature Photonics}\ }\textbf
  {\bibinfo {volume} {12}},\ \bibinfo {pages} {613--620} (\bibinfo {year}
  {2018})}\BibitemShut {NoStop}%
\bibitem [{\citenamefont {Kim}\ \emph {et~al.}(2015)\citenamefont {Kim},
  \citenamefont {Kuzyk}, \citenamefont {Han}, \citenamefont {Wang},\ and\
  \citenamefont {Bahl}}]{Kim2015}%
  \BibitemOpen
  \bibfield  {author} {\bibinfo {author} {\bibfnamefont {JunHwan}\ \bibnamefont
  {Kim}}, \bibinfo {author} {\bibfnamefont {Mark~C.}\ \bibnamefont {Kuzyk}},
  \bibinfo {author} {\bibfnamefont {Kewen}\ \bibnamefont {Han}}, \bibinfo
  {author} {\bibfnamefont {Hailin}\ \bibnamefont {Wang}}, \ and\ \bibinfo
  {author} {\bibfnamefont {Gaurav}\ \bibnamefont {Bahl}},\ }\bibfield  {title}
  {\enquote {\bibinfo {title} {{Non-reciprocal Brillouin scattering induced
  transparency}},}\ }\href {\doibase 10.1038/nphys3236} {\bibfield  {journal}
  {\bibinfo  {journal} {Nature Physics}\ }\textbf {\bibinfo {volume} {11}},\
  \bibinfo {pages} {275--280} (\bibinfo {year} {2015})}\BibitemShut {NoStop}%
\bibitem [{\citenamefont {Dong}\ \emph {et~al.}(2015)\citenamefont {Dong},
  \citenamefont {Shen}, \citenamefont {Zou}, \citenamefont {Zhang},
  \citenamefont {Fu},\ and\ \citenamefont {Guo}}]{Dong2015}%
  \BibitemOpen
  \bibfield  {author} {\bibinfo {author} {\bibfnamefont {Chun-Hua}\
  \bibnamefont {Dong}}, \bibinfo {author} {\bibfnamefont {Zhen}\ \bibnamefont
  {Shen}}, \bibinfo {author} {\bibfnamefont {Chang-Ling}\ \bibnamefont {Zou}},
  \bibinfo {author} {\bibfnamefont {Yan-Lei}\ \bibnamefont {Zhang}}, \bibinfo
  {author} {\bibfnamefont {Wei}\ \bibnamefont {Fu}}, \ and\ \bibinfo {author}
  {\bibfnamefont {Guang-Can}\ \bibnamefont {Guo}},\ }\bibfield  {title}
  {\enquote {\bibinfo {title} {{Brillouin-scattering-induced transparency and
  non-reciprocal light storage}},}\ }\href {\doibase 10.1038/ncomms7193}
  {\bibfield  {journal} {\bibinfo  {journal} {Nature Communications}\ }\textbf
  {\bibinfo {volume} {6}},\ \bibinfo {pages} {6193} (\bibinfo {year}
  {2015})}\BibitemShut {NoStop}%
\bibitem [{\citenamefont {Kim}\ \emph {et~al.}(2017)\citenamefont {Kim},
  \citenamefont {Kim},\ and\ \citenamefont {Bahl}}]{Kim2017a}%
  \BibitemOpen
  \bibfield  {author} {\bibinfo {author} {\bibfnamefont {Jun~Hwan}\
  \bibnamefont {Kim}}, \bibinfo {author} {\bibfnamefont {Seunghwi}\
  \bibnamefont {Kim}}, \ and\ \bibinfo {author} {\bibfnamefont {Gaurav}\
  \bibnamefont {Bahl}},\ }\bibfield  {title} {\enquote {\bibinfo {title}
  {{Complete linear optical isolation at the microscale with ultralow loss}},}\
  }\href {\doibase 10.1038/s41598-017-01494-w} {\bibfield  {journal} {\bibinfo
  {journal} {Scientific Reports}\ }\textbf {\bibinfo {volume} {7}},\ \bibinfo
  {pages} {1--9} (\bibinfo {year} {2017})}\BibitemShut {NoStop}%
\bibitem [{\citenamefont {Stiller}\ \emph
  {et~al.}(2019{\natexlab{a}})\citenamefont {Stiller}, \citenamefont
  {Merklein}, \citenamefont {Vu}, \citenamefont {Ma}, \citenamefont {Madden},
  \citenamefont {Poulton},\ and\ \citenamefont {Eggleton}}]{Stiller2019}%
  \BibitemOpen
  \bibfield  {author} {\bibinfo {author} {\bibfnamefont {Birgit}\ \bibnamefont
  {Stiller}}, \bibinfo {author} {\bibfnamefont {Moritz}\ \bibnamefont
  {Merklein}}, \bibinfo {author} {\bibfnamefont {Khu}\ \bibnamefont {Vu}},
  \bibinfo {author} {\bibfnamefont {Pan}\ \bibnamefont {Ma}}, \bibinfo {author}
  {\bibfnamefont {Stephen~J.}\ \bibnamefont {Madden}}, \bibinfo {author}
  {\bibfnamefont {Christopher~G.}\ \bibnamefont {Poulton}}, \ and\ \bibinfo
  {author} {\bibfnamefont {Benjamin~J.}\ \bibnamefont {Eggleton}},\ }\bibfield
  {title} {\enquote {\bibinfo {title} {{Cross talk-free coherent
  multi-wavelength Brillouin interaction}},}\ }\href {\doibase
  10.1063/1.5087180} {\bibfield  {journal} {\bibinfo  {journal} {APL
  Photonics}\ }\textbf {\bibinfo {volume} {4}},\ \bibinfo {pages} {040802}
  (\bibinfo {year} {2019}{\natexlab{a}})}\BibitemShut {NoStop}%
\bibitem [{\citenamefont {Merklein}\ \emph {et~al.}(2017)\citenamefont
  {Merklein}, \citenamefont {Stiller}, \citenamefont {Vu}, \citenamefont
  {Madden},\ and\ \citenamefont {Eggleton}}]{Merklein2017}%
  \BibitemOpen
  \bibfield  {author} {\bibinfo {author} {\bibfnamefont {Moritz}\ \bibnamefont
  {Merklein}}, \bibinfo {author} {\bibfnamefont {Birgit}\ \bibnamefont
  {Stiller}}, \bibinfo {author} {\bibfnamefont {Khu}\ \bibnamefont {Vu}},
  \bibinfo {author} {\bibfnamefont {Stephen~J.}\ \bibnamefont {Madden}}, \ and\
  \bibinfo {author} {\bibfnamefont {Benjamin~J.}\ \bibnamefont {Eggleton}},\
  }\bibfield  {title} {\enquote {\bibinfo {title} {{A chip-integrated coherent
  photonic-phononic memory}},}\ }\href {\doibase 10.1038/s41467-017-00717-y}
  {\bibfield  {journal} {\bibinfo  {journal} {Nature Communications}\ }\textbf
  {\bibinfo {volume} {8}},\ \bibinfo {pages} {574} (\bibinfo {year}
  {2017})}\BibitemShut {NoStop}%
\bibitem [{\citenamefont {Zhu}\ \emph {et~al.}(2007)\citenamefont {Zhu},
  \citenamefont {Gauthier},\ and\ \citenamefont {Boyd}}]{Zhu2007}%
  \BibitemOpen
  \bibfield  {author} {\bibinfo {author} {\bibfnamefont {Zhaoming}\
  \bibnamefont {Zhu}}, \bibinfo {author} {\bibfnamefont {Daniel~J}\
  \bibnamefont {Gauthier}}, \ and\ \bibinfo {author} {\bibfnamefont {Robert~W}\
  \bibnamefont {Boyd}},\ }\bibfield  {title} {\enquote {\bibinfo {title}
  {{Stored light in an optical fiber via stimulated Brillouin scattering.}}}\
  }\href {\doibase 10.1126/science.1149066} {\bibfield  {journal} {\bibinfo
  {journal} {Science}\ }\textbf {\bibinfo {volume} {318}},\ \bibinfo {pages}
  {1748--50} (\bibinfo {year} {2007})}\BibitemShut {NoStop}%
\bibitem [{\citenamefont {Merklein}\ \emph {et~al.}(2018)\citenamefont
  {Merklein}, \citenamefont {Stiller},\ and\ \citenamefont
  {Eggleton}}]{Merklein2018}%
  \BibitemOpen
  \bibfield  {author} {\bibinfo {author} {\bibfnamefont {M}~\bibnamefont
  {Merklein}}, \bibinfo {author} {\bibfnamefont {B}~\bibnamefont {Stiller}}, \
  and\ \bibinfo {author} {\bibfnamefont {B~J}\ \bibnamefont {Eggleton}},\
  }\bibfield  {title} {\enquote {\bibinfo {title} {{Brillouin-based light
  storage and delay techniques}},}\ }\href {\doibase 10.1088/2040-8986/aad081}
  {\bibfield  {journal} {\bibinfo  {journal} {Journal of Optics}\ }\textbf
  {\bibinfo {volume} {20}},\ \bibinfo {pages} {083003} (\bibinfo {year}
  {2018})}\BibitemShut {NoStop}%
\bibitem [{\citenamefont {Poulton}\ \emph {et~al.}(2013)\citenamefont
  {Poulton}, \citenamefont {Pant},\ and\ \citenamefont
  {Eggleton}}]{Poulton2013a}%
  \BibitemOpen
  \bibfield  {author} {\bibinfo {author} {\bibfnamefont {Christopher~G.}\
  \bibnamefont {Poulton}}, \bibinfo {author} {\bibfnamefont {Ravi}\
  \bibnamefont {Pant}}, \ and\ \bibinfo {author} {\bibfnamefont {Benjamin~J.}\
  \bibnamefont {Eggleton}},\ }\bibfield  {title} {\enquote {\bibinfo {title}
  {{Acoustic confinement and stimulated Brillouin scattering in integrated
  optical waveguides}},}\ }\href {\doibase 10.1364/JOSAB.30.002657} {\bibfield
  {journal} {\bibinfo  {journal} {Journal of the Optical Society of America B}\
  }\textbf {\bibinfo {volume} {30}},\ \bibinfo {pages} {2657--2664} (\bibinfo
  {year} {2013})}\BibitemShut {NoStop}%
\bibitem [{\citenamefont {Fiore}\ \emph {et~al.}(2011)\citenamefont {Fiore},
  \citenamefont {Yang}, \citenamefont {Kuzyk}, \citenamefont {Barbour},
  \citenamefont {Tian},\ and\ \citenamefont {Wang}}]{Fiore2011a}%
  \BibitemOpen
  \bibfield  {author} {\bibinfo {author} {\bibfnamefont {Victor}\ \bibnamefont
  {Fiore}}, \bibinfo {author} {\bibfnamefont {Yong}\ \bibnamefont {Yang}},
  \bibinfo {author} {\bibfnamefont {Mark~C.}\ \bibnamefont {Kuzyk}}, \bibinfo
  {author} {\bibfnamefont {Russell}\ \bibnamefont {Barbour}}, \bibinfo {author}
  {\bibfnamefont {Lin}\ \bibnamefont {Tian}}, \ and\ \bibinfo {author}
  {\bibfnamefont {Hailin}\ \bibnamefont {Wang}},\ }\bibfield  {title} {\enquote
  {\bibinfo {title} {{Storing optical information as a mechanical excitation in
  a silica optomechanical resonator}},}\ }\href {\doibase
  10.1103/PhysRevLett.107.133601} {\bibfield  {journal} {\bibinfo  {journal}
  {Physical Review Letters}\ }\textbf {\bibinfo {volume} {107}},\ \bibinfo
  {pages} {1--5} (\bibinfo {year} {2011})}\BibitemShut {NoStop}%
\bibitem [{\citenamefont {Fiore}\ \emph {et~al.}(2013)\citenamefont {Fiore},
  \citenamefont {Dong}, \citenamefont {Kuzyk},\ and\ \citenamefont
  {Wang}}]{Fiore2013}%
  \BibitemOpen
  \bibfield  {author} {\bibinfo {author} {\bibfnamefont {Victor}\ \bibnamefont
  {Fiore}}, \bibinfo {author} {\bibfnamefont {Chunhua}\ \bibnamefont {Dong}},
  \bibinfo {author} {\bibfnamefont {Mark~C.}\ \bibnamefont {Kuzyk}}, \ and\
  \bibinfo {author} {\bibfnamefont {Hailin}\ \bibnamefont {Wang}},\ }\bibfield
  {title} {\enquote {\bibinfo {title} {{Optomechanical light storage in a
  silica microresonator}},}\ }\href {\doibase 10.1103/PhysRevA.87.023812}
  {\bibfield  {journal} {\bibinfo  {journal} {Physical Review A}\ }\textbf
  {\bibinfo {volume} {87}},\ \bibinfo {pages} {023812} (\bibinfo {year}
  {2013})}\BibitemShut {NoStop}%
\bibitem [{\citenamefont {Okawachi}\ \emph {et~al.}(2005)\citenamefont
  {Okawachi}, \citenamefont {Bigelow}, \citenamefont {Sharping}, \citenamefont
  {Zhu}, \citenamefont {Schweinsberg}, \citenamefont {Gauthier}, \citenamefont
  {Boyd},\ and\ \citenamefont {Gaeta}}]{Okawachi2005}%
  \BibitemOpen
  \bibfield  {author} {\bibinfo {author} {\bibfnamefont {Yoshitomo}\
  \bibnamefont {Okawachi}}, \bibinfo {author} {\bibfnamefont {Matthew}\
  \bibnamefont {Bigelow}}, \bibinfo {author} {\bibfnamefont {Jay}\ \bibnamefont
  {Sharping}}, \bibinfo {author} {\bibfnamefont {Zhaoming}\ \bibnamefont
  {Zhu}}, \bibinfo {author} {\bibfnamefont {Aaron}\ \bibnamefont
  {Schweinsberg}}, \bibinfo {author} {\bibfnamefont {Daniel}\ \bibnamefont
  {Gauthier}}, \bibinfo {author} {\bibfnamefont {Robert}\ \bibnamefont {Boyd}},
  \ and\ \bibinfo {author} {\bibfnamefont {Alexander}\ \bibnamefont {Gaeta}},\
  }\bibfield  {title} {\enquote {\bibinfo {title} {{Tunable All-Optical Delays
  via Brillouin Slow Light in an Optical Fiber}},}\ }\href {\doibase
  10.1103/PhysRevLett.94.153902} {\bibfield  {journal} {\bibinfo  {journal}
  {Physical Review Letters}\ }\textbf {\bibinfo {volume} {94}},\ \bibinfo
  {pages} {153902} (\bibinfo {year} {2005})}\BibitemShut {NoStop}%
\bibitem [{\citenamefont {Song}\ \emph {et~al.}(2005)\citenamefont {Song},
  \citenamefont {Herr{\'{a}}ez},\ and\ \citenamefont
  {Th{\'{e}}venaz}}]{Song2005}%
  \BibitemOpen
  \bibfield  {author} {\bibinfo {author} {\bibfnamefont {Kwang~Yong}\
  \bibnamefont {Song}}, \bibinfo {author} {\bibfnamefont {Miguel}\ \bibnamefont
  {Herr{\'{a}}ez}}, \ and\ \bibinfo {author} {\bibfnamefont {Luc}\ \bibnamefont
  {Th{\'{e}}venaz}},\ }\bibfield  {title} {\enquote {\bibinfo {title}
  {{Observation of pulse delaying and advancement in optical fibers using
  stimulated Brillouin scattering.}}}\ }\href {\doibase 10.1364/OPEX.13.000082}
  {\bibfield  {journal} {\bibinfo  {journal} {Optics Express}\ }\textbf
  {\bibinfo {volume} {13}},\ \bibinfo {pages} {82--88} (\bibinfo {year}
  {2005})}\BibitemShut {NoStop}%
\bibitem [{\citenamefont {Song}\ \emph {et~al.}(2009)\citenamefont {Song},
  \citenamefont {Lee},\ and\ \citenamefont {Lee}}]{Song2009}%
  \BibitemOpen
  \bibfield  {author} {\bibinfo {author} {\bibfnamefont {Kwang~Yong}\
  \bibnamefont {Song}}, \bibinfo {author} {\bibfnamefont {Kwanil}\ \bibnamefont
  {Lee}}, \ and\ \bibinfo {author} {\bibfnamefont {Sang~Bae}\ \bibnamefont
  {Lee}},\ }\bibfield  {title} {\enquote {\bibinfo {title} {{Tunable optical
  delays based on Brillouin dynamic grating in optical fibers.}}}\ }\href
  {http://www.ncbi.nlm.nih.gov/pubmed/19506688} {\bibfield  {journal} {\bibinfo
   {journal} {Optics Express}\ }\textbf {\bibinfo {volume} {17}},\ \bibinfo
  {pages} {10344--9} (\bibinfo {year} {2009})}\BibitemShut {NoStop}%
\bibitem [{\citenamefont {Preussler}\ \emph {et~al.}(2009)\citenamefont
  {Preussler}, \citenamefont {Jamshidi}, \citenamefont {Wiatrek}, \citenamefont
  {Henker}, \citenamefont {Bunge},\ and\ \citenamefont
  {Schneider}}]{Preussler2009}%
  \BibitemOpen
  \bibfield  {author} {\bibinfo {author} {\bibfnamefont {Stefan}\ \bibnamefont
  {Preussler}}, \bibinfo {author} {\bibfnamefont {Kambiz}\ \bibnamefont
  {Jamshidi}}, \bibinfo {author} {\bibfnamefont {Andrzej}\ \bibnamefont
  {Wiatrek}}, \bibinfo {author} {\bibfnamefont {Ronny}\ \bibnamefont {Henker}},
  \bibinfo {author} {\bibfnamefont {Christian-Alexander}\ \bibnamefont
  {Bunge}}, \ and\ \bibinfo {author} {\bibfnamefont {Thomas}\ \bibnamefont
  {Schneider}},\ }\bibfield  {title} {\enquote {\bibinfo {title}
  {{Quasi-Light-Storage based on time-frequency coherence.}}}\ }\href {\doibase
  10.1364/OE.17.015790} {\bibfield  {journal} {\bibinfo  {journal} {Optics
  Express}\ }\textbf {\bibinfo {volume} {17}},\ \bibinfo {pages} {15790--15798}
  (\bibinfo {year} {2009})}\BibitemShut {NoStop}%
\bibitem [{\citenamefont {Chin}\ and\ \citenamefont
  {Th{\'{e}}venaz}(2012)}]{Chin2012}%
  \BibitemOpen
  \bibfield  {author} {\bibinfo {author} {\bibfnamefont {S.}~\bibnamefont
  {Chin}}\ and\ \bibinfo {author} {\bibfnamefont {L.}~\bibnamefont
  {Th{\'{e}}venaz}},\ }\bibfield  {title} {\enquote {\bibinfo {title} {{Tunable
  photonic delay lines in optical fibers}},}\ }\href {\doibase
  10.1002/lpor.201100038} {\bibfield  {journal} {\bibinfo  {journal} {Laser
  {\&} Photonics Reviews}\ }\textbf {\bibinfo {volume} {6}},\ \bibinfo {pages}
  {724--738} (\bibinfo {year} {2012})}\BibitemShut {NoStop}%
\bibitem [{\citenamefont {Stiller}\ \emph
  {et~al.}(2019{\natexlab{b}})\citenamefont {Stiller}, \citenamefont
  {Merklein}, \citenamefont {Wolff}, \citenamefont {Vu}, \citenamefont {Ma},
  \citenamefont {Madden},\ and\ \citenamefont {Eggleton}}]{Stiller2019a}%
  \BibitemOpen
  \bibfield  {author} {\bibinfo {author} {\bibfnamefont {Birgit}\ \bibnamefont
  {Stiller}}, \bibinfo {author} {\bibfnamefont {Moritz}\ \bibnamefont
  {Merklein}}, \bibinfo {author} {\bibfnamefont {Christian}\ \bibnamefont
  {Wolff}}, \bibinfo {author} {\bibfnamefont {Khu}\ \bibnamefont {Vu}},
  \bibinfo {author} {\bibfnamefont {Pan}\ \bibnamefont {Ma}}, \bibinfo {author}
  {\bibfnamefont {Stephen~J}\ \bibnamefont {Madden}}, \ and\ \bibinfo {author}
  {\bibfnamefont {Benjamin~J}\ \bibnamefont {Eggleton}},\ }\bibfield  {title}
  {\enquote {\bibinfo {title} {{Coherently refreshed acoustic phonons for
  extended light storage}},}\ }\href {http://arxiv.org/abs/1904.13167}
  {\bibfield  {journal} {\bibinfo  {journal} {arXiv: 1904.13167}\ ,\ \bibinfo
  {pages} {1--6}} (\bibinfo {year} {2019}{\natexlab{b}})}\BibitemShut {NoStop}%
\bibitem [{\citenamefont {Madden}\ \emph {et~al.}(2007)\citenamefont {Madden},
  \citenamefont {Choi}, \citenamefont {Bulla}, \citenamefont {Rode},
  \citenamefont {Luther-Davies}, \citenamefont {Ta'eed}, \citenamefont
  {Pelusi},\ and\ \citenamefont {Eggleton}}]{Madden2007}%
  \BibitemOpen
  \bibfield  {author} {\bibinfo {author} {\bibfnamefont {S~J}\ \bibnamefont
  {Madden}}, \bibinfo {author} {\bibfnamefont {D-Y}\ \bibnamefont {Choi}},
  \bibinfo {author} {\bibfnamefont {D~a}\ \bibnamefont {Bulla}}, \bibinfo
  {author} {\bibfnamefont {a~V}\ \bibnamefont {Rode}}, \bibinfo {author}
  {\bibfnamefont {B}~\bibnamefont {Luther-Davies}}, \bibinfo {author}
  {\bibfnamefont {V~G}\ \bibnamefont {Ta'eed}}, \bibinfo {author}
  {\bibfnamefont {M~D}\ \bibnamefont {Pelusi}}, \ and\ \bibinfo {author}
  {\bibfnamefont {B~J}\ \bibnamefont {Eggleton}},\ }\bibfield  {title}
  {\enquote {\bibinfo {title} {{Long, low loss etched As(2)S(3) chalcogenide
  waveguides for all-optical signal regeneration.}}}\ }\href
  {http://www.ncbi.nlm.nih.gov/pubmed/19550720} {\bibfield  {journal} {\bibinfo
   {journal} {Optics Express}\ }\textbf {\bibinfo {volume} {15}},\ \bibinfo
  {pages} {14414--14421} (\bibinfo {year} {2007})}\BibitemShut {NoStop}%
\bibitem [{\citenamefont {Pant}\ \emph {et~al.}(2011)\citenamefont {Pant},
  \citenamefont {Poulton}, \citenamefont {Choi}, \citenamefont {Mcfarlane},
  \citenamefont {Hile}, \citenamefont {Li}, \citenamefont {Thevenaz},
  \citenamefont {Luther-Davies}, \citenamefont {Madden},\ and\ \citenamefont
  {Eggleton}}]{Pant2011}%
  \BibitemOpen
  \bibfield  {author} {\bibinfo {author} {\bibfnamefont {Ravi}\ \bibnamefont
  {Pant}}, \bibinfo {author} {\bibfnamefont {Christopher~G.}\ \bibnamefont
  {Poulton}}, \bibinfo {author} {\bibfnamefont {Duk-Yong}\ \bibnamefont
  {Choi}}, \bibinfo {author} {\bibfnamefont {Hannah}\ \bibnamefont
  {Mcfarlane}}, \bibinfo {author} {\bibfnamefont {Samuel}\ \bibnamefont
  {Hile}}, \bibinfo {author} {\bibfnamefont {Enbang}\ \bibnamefont {Li}},
  \bibinfo {author} {\bibfnamefont {Luc}\ \bibnamefont {Thevenaz}}, \bibinfo
  {author} {\bibfnamefont {Barry}\ \bibnamefont {Luther-Davies}}, \bibinfo
  {author} {\bibfnamefont {Stephen~J.}\ \bibnamefont {Madden}}, \ and\ \bibinfo
  {author} {\bibfnamefont {Benjamin~J.}\ \bibnamefont {Eggleton}},\ }\bibfield
  {title} {\enquote {\bibinfo {title} {{On-chip stimulated Brillouin
  scattering}},}\ }\href {\doibase 10.1364/OE.19.008285} {\bibfield  {journal}
  {\bibinfo  {journal} {Optics Express}\ }\textbf {\bibinfo {volume} {19}},\
  \bibinfo {pages} {8285--8290} (\bibinfo {year} {2011})}\BibitemShut {NoStop}%
\bibitem [{\citenamefont {Zarifi}\ \emph {et~al.}(2018)\citenamefont {Zarifi},
  \citenamefont {Stiller}, \citenamefont {Merklein}, \citenamefont {Li},
  \citenamefont {Vu}, \citenamefont {Choi}, \citenamefont {Ma}, \citenamefont
  {Madden},\ and\ \citenamefont {Eggleton}}]{Zarifi2017}%
  \BibitemOpen
  \bibfield  {author} {\bibinfo {author} {\bibfnamefont {Atiyeh}\ \bibnamefont
  {Zarifi}}, \bibinfo {author} {\bibfnamefont {Birgit}\ \bibnamefont
  {Stiller}}, \bibinfo {author} {\bibfnamefont {Moritz}\ \bibnamefont
  {Merklein}}, \bibinfo {author} {\bibfnamefont {Neuton}\ \bibnamefont {Li}},
  \bibinfo {author} {\bibfnamefont {Khu}\ \bibnamefont {Vu}}, \bibinfo {author}
  {\bibfnamefont {Duk-Yong}\ \bibnamefont {Choi}}, \bibinfo {author}
  {\bibfnamefont {Pan}\ \bibnamefont {Ma}}, \bibinfo {author} {\bibfnamefont
  {Stephen~J.}\ \bibnamefont {Madden}}, \ and\ \bibinfo {author} {\bibfnamefont
  {Benjamin~J.}\ \bibnamefont {Eggleton}},\ }\bibfield  {title} {\enquote
  {\bibinfo {title} {{Highly localized distributed Brillouin scattering
  response in a photonic integrated circuit}},}\ }\href {\doibase
  10.1063/1.5000108} {\bibfield  {journal} {\bibinfo  {journal} {APL
  Photonics}\ }\textbf {\bibinfo {volume} {3}},\ \bibinfo {pages} {036101}
  (\bibinfo {year} {2018})}\BibitemShut {NoStop}%
\end{thebibliography}%
%
%
\end{document}